\documentclass[fleqn,usenatbib]{mnras}

\usepackage{newtxtext,newtxmath}

\usepackage[T1]{fontenc}

\usepackage{xcolor}

\DeclareRobustCommand{\VAN}[3]{#2}
\let\VANthebibliography\thebibliography
\def\thebibliography{\DeclareRobustCommand{\VAN}[3]{##3}\VANthebibliography}

\usepackage{graphicx}	
\usepackage{amsmath}	
\usepackage{bm}

\graphicspath{{./}}
\newcommand{\tref}[1]{Table~\ref{#1}}
\newcommand{\fref}[1]{Fig.~\ref{#1}}
\newcommand{\eref}[1]{equation~\eqref{#1}}
\newcommand{\sref}[1]{Section~\ref{#1}}
\newcommand{\vbc}{v_{\rm bc}}
\newcommand{\msolh}{h^{-1}~\mathrm{M}_\odot}
\newcommand{\msol}{\mathrm{M}_\odot}
\newcommand{\mpch}{h^{-1}~\mathrm{Mpc}}
\newcommand{\impch}{h~\mathrm{Mpc}^{-1}}
\newcommand{\kpch}{h^{-1}~\mathrm{kpc}}
\newcommand{\pch}{h^{-1}~\mathrm{pc}}
\newcommand{\kms}{\mathrm{km}~\mathrm{s}^{-1}}
\newcommand{\myr}{\mathrm{Myr}}
\newcommand{\om}{\Omega_{\rm m}}
\newcommand{\ob}{\Omega_{\rm b}}
\newcommand{\ol}{\Omega_{\Lambda}}

\newcommand{\fb}{f_{\rm b}}

\newcommand{\novbc}{$\text{no}~\vbc$}
\newcommand{\biased}{\mbox{$\vbc$--rec}}
\newcommand{\unbiased}{\mbox{$\vbc$--ini}}
\newcommand{\ramses}{{\sc ramses}}

\newcommand{\com}[2]{#2}
\newcommand{\rev}[2]{#2}
\newcommand{\aut}[2]{#2}

\title[Baryon drift]{Relative baryon-dark matter velocities in
  cosmological zoom simulations}

\author[L. Conaboy et al.]{Luke Conaboy$^{1,2}$\thanks{E-mail: luke.conaboy@nottingham.ac.uk},
Ilian T. Iliev$^{1}$,
Anastasia Fialkov$^{3,4}$,
Keri L. Dixon$^{5,6}$ and
David Sullivan$^{1}$
\\
$^{1}$Astronomy Centre, Department of Physics \& Astronomy, University of Sussex, Brighton, BN1 9QH, UK\\
$^{2}$School of Physics and Astronomy, University of Nottingham, University Park, Nottingham, NG7 2RD, UK\\
$^{3}$Institute of Astronomy, University of Cambridge, Madingley Road, Cambridge CB3 0HA, UK\\
$^{4}$Kavli Institute for Cosmology, University of Cambridge, Madingley Road, Cambridge CB3 0HA, UK\\
$^{5}$New York University Abu Dhabi, PO Box 129188, Saadiyat Island,
Abu Dhabi, United Arab Emirates\\
$^{6}$Center for Astro, Particle and Planetary Physics (CAP$^{\text{3}}$), New York University Abu Dhabi}

\date{Accepted 2023 August 25. Received 2023 August 25; in original form 2022 July 22}

\pubyear{2023}

\begin{document}
\label{firstpage}
\pagerange{\pageref{firstpage}--\pageref{lastpage}}
\maketitle

\begin{abstract}
  Supersonic relative motion between baryons and dark matter due to
  the decoupling of baryons from the primordial plasma after
  recombination affects the growth of the first small-scale
  structures. Large box sizes (greater than a few hundred Mpc) are
  required to sample the full range of scales pertinent to the
  relative velocity, while the effect of the relative velocity is
  strongest on small scales (less than a few hundred kpc). This
  separation of scales naturally lends itself to the use of `zoom'
  simulations, and here we present our methodology to
  self-consistently \rev{A}{incorporate the relative velocity in zoom
    simulations, including its cumulative effect from recombination
    through to the start time of the simulation}. \aut{8}{We apply our
    methodology to a large-scale cosmological zoom simulation, finding
    that the inclusion of relative velocities suppresses the halo
    baryon fraction by $46$--$23$ per cent between $z=13.6$ and
    $11.2$, in qualitative agreement \rev{}{with} previous works. In
    addition, we find that including the relative velocity delays the
    formation of star particles by $\sim20~\myr$ on average (of the order
    of the lifetime of a $\sim9~\msol$ Population III star) and
    suppresses the final stellar mass by as much as $79$ per cent at
    $z=11.2$.}
\end{abstract}





\section{Introduction}
\label{sec:int}

The cosmic microwave background (CMB) radiation carries an image of
the Universe at the moment of recombination, when the first neutral
atoms formed at $z_{\mathrm{rec}} \approx 1089$. Prior to recombination,
photons and baryons were tightly coupled and oscillated as a single
plasma \com{1.2}{until they decoupled at
  $z_{\mathrm{dec}}\approx 1020$}. These oscillations, referred to as the
baryon acoustic oscillations (BAO), are observed today as fluctuations
of the CMB temperature
\citep[e.g.][]{planckcollaboration2020}. \com{1.3}{Acoustic
  oscillations in the \com{1.1}{baryons'} velocity at the moment of
  their decoupling lead to clumping in the distribution of baryons at
  later times, resulting in over and underdense regions.}  The
initially tiny perturbations grew under the effect of gravity and are
detected today in the distribution of galaxies on the largest
cosmological scales \citep[e.g.][]{alam2017}.

\com{1.4}{Not only did the plasma oscillations lead to the BAO
  features in the post-recombination distribution of baryons, they
  also impacted the baryons' velocities \citep{sunyaev1970}.}
\citet{tseliakhovich2010} were the first to point out that the BAO
pattern is imprinted in the magnitude of the relative velocity between
baryons and dark matter, because the latter was not coupled to the
primordial plasma at the time of recombination. At decoupling, the
relative velocity had a root-mean-square (RMS) of
$\langle\vbc^2\rangle^{\frac{1}{2}}\approx30~\kms$, or
$\sim 10^{-4}\,c$ with $c$ the speed of light. \com{1.5}{There is a vast
  separation of scales relevant to the relative velocity. Over scales
  smaller than a few Mpc (the coherence scale), the relative velocity
  is roughly constant; however, box sizes greater than a few hundred
  Mpc are required to properly sample the relative velocity \citep[see
  \fref{fig:rms-vbc} \rev{}{of this work} and also fig.~1
  in][]{tseliakhovich2010}.}  At recombination the sound speed of the
baryonic fluid dropped from being relativistic, $\sim c/\sqrt{3}$, to the
thermal velocities of hydrogen atoms, $\sim 2 \times 10^{-5}c$, which is much
smaller than the RMS value of $\vbc$. Therefore, on average, at
decoupling baryons were travelling with supersonic velocities relative
to the underlying potential wells generated by dark matter haloes
\citep{tseliakhovich2010}. The relative velocity remained supersonic
all the way down to $z\sim 15$, sourcing shocks and entropy generation
\citep{tseliakhovich2010, oleary2012}.  The amplitude of the velocity
field decayed with time as $(1+z)$, and thus the effect weakened as
the Universe expanded. \com{1.6}{For instance, the signature of $\vbc$
  in the low-$z$ power spectrum of BOSS galaxies \citep{yoo2013,
    beutler2017} and the three-point correlation function of BOSS
  CMASS galaxies was found to be negligible \citep{slepian2015,
    slepian2018}.}

In the post-recombination Universe, growth of structure on large
cosmological scales is generally described by linear perturbation
theory, which follows the evolution of density and velocity fields to
the leading order in perturbations. Despite being formally
second-order contributions, terms that involve the supersonic relative
velocity can actually be as large as the first order terms. Moreover,
on scales below the coherence scale, $\vbc$ is
position-independent and the second-order terms become effectively
linear \citep{tseliakhovich2010}. Using such a `quasi-linear'
approach, analytical methods were employed to explore implications of
$\vbc$ in the cosmological context. Supersonic relative
velocities modulate the abundance of minihaloes and their gas content
on the BAO scale \citep[e.g.][]{tseliakhovich2010, tseliakhovich2011,
  fialkov2012, ahn2016, ahn2018}, affecting fluctuations of the 21-cm
signal of neutral hydrogen \aut{5}{\citep[e.g.][]{dalal2010, mcquinn2012,
  visbal2012, fialkov2013, cohen2016, fialkov2018, munoz2019, cain2020,
  munoz2022, long2022}}.

Numerical simulations were used to explore non-linear effects of
$v_{\rm bc}$ on scales well below its coherence scale. Such
simulations typically employ boxes of several comoving Mpc or less and
assume a position-independent velocity field. These simulations
demonstrated that $v_{\rm bc}$ suppresses formation of small dark
matter haloes \citep{naoz2012, naoz2013, oleary2012}, induces shocks
\citep{oleary2012}, affects the formation of first stars
\citep[e.g.][]{maio2011, stacy2011, greif2011, schauer2019,
  schauer2021a} and black holes \citep[e.g.][]{hirano2017,
  schauer2017}, and \com{1.7}{may even \rev{A}{influence}} shaping
globular clusters \aut{5}{\citep{naoz2014, chiou2018, chiou2019,
    chiou2021, druschke2020, lake2021}}.

Finally, a hybrid approach was used to incorporate the non-linear
effects into the large-scale cosmological picture by tiling regions of
fixed $v_{\rm bc}$ together \citep[e.g.][]{visbal2012,
  fialkov2013}. In such studies, the distribution of $v_{\rm bc}$ on
scales larger than the `pixel' size was generated from the
corresponding density field using the continuity equation, while star
formation in each `pixel' was calibrated to numerical simulations
\citep{maio2011, stacy2011, greif2011, naoz2012, naoz2013}. This
method was applied to the 21-cm signal of neutral hydrogen revealing
enhanced BAO patterns \citep{visbal2012, fialkov2013}.

To fully capture the non-linear effect of $v_{\rm bc}$ in the
cosmological context, it is necessary to properly include the velocity
effect in the initial conditions of $N$-body and hydrodynamical
simulations. Such a task would require an accurate non-linear
treatment of dark matter, baryons, and radiation, starting at
$z_{\mathrm{rec}}$ and following the growth of structure all the way
down to the lowest simulated redshift. This treatment is not possible
due to the large dynamical range: the amplitude of density
fluctuations at recombination is smaller than the precision of many
commonly used integration schemes. Today, state-of-the-art numerical
simulations are typically initialised at $z_{\rm ini}\sim 200$ with
fields that do not \com{1.8}{incorporate the effect of the relative velocity}.
 
Recently, \citet{ahn2018} introduced a new cosmological initial
condition generator {\sc bccomics}, based on a code that solves the
quasi-linear equations between $z_{\rm rec}$ and $z_{\rm ini}$ for
fixed values of large-scale density, $\delta$, and $v_{\rm bc}$ at
decoupling \citep{ahn2016}. Next, the solver is applied to a larger
cosmic volume divided in regions of fixed $\delta$ and $v_{\rm bc}$. The
code simulates the growth of small-scale structure inside density
peaks and voids, by treating each patch of fixed $\delta$ and
$v_{\rm bc}$ as a separate universe. \com{1.10}{They do this because
  the evolution equations from \citet{tseliakhovich2010} are only
  valid at mean density, and \citet{ahn2016} found that structure
  formation was boosted (suppressed) in overdense (underdense)
  patches.} \rev{}{They estimated the} effect of $v_{\rm bc}$ and
$\delta$ on the abundance of small haloes and, in some
cases, \rev{}{reported} qualitative disagreement with prior works.

Here we take an independent approach and develop a new initial
conditions generator. \rev{A}{Our approach differs from that of {\sc
    bccomics} as, instead of treating small patches as separate
  universes, we follow small patches embedded in a larger cosmological
  volume through the use of zoom simulations. We compensate for the
  lacking effect of $v_{\rm bc}$ between $z_{\rm rec}$ and
  $z_{\rm ini}= 200$ through the use of a bias factor, described in
  \sref{sec:bias}.} After this compensation, the effect of $\vbc$ from
$z_{\rm ini}= 200$ is naturally included by the simulation, through
initialising the simulations with separate transfer functions for the
baryon and dark matter velocities. \com{1.10}{Unlike \citet{ahn2016},
  we do not account for the large-scale density environment, since our
  simulation focuses on a patch with very near to mean density and so
  we can safely ignore the impact on structure formation due to
  non-zero overdensity.} Our methodology employs the widely used code
{\sc music} \citep{hahn2011} to generate high-resolution `zoom'
initial conditions \citep{bertschinger2001} in large cosmic volumes
and by design is well-matched to AMR simulations. We demonstrate the
performance by generating initial conditions in a $400~\mpch$ box
before extracting a $100~\mpch$ subbox, which is used to run a
simulation from $z_{\rm ini}=200$ to the final redshift of $11.2$ with
the AMR code {\sc ramses} \citep{teyssier2002}. We explore the
performance by computing the effects of $v_{\rm bc}$ on the number of
haloes formed, baryon fraction, and star formation.

\rev{A}{}The paper is organised as follows: in \sref{sec:th}, we recap the
theoretical background and discuss why large simulation box sizes are
needed; in \sref{sec:method}, we discuss the simulation setup and our
methodology for incorporating the effect of the $\vbc$ through a
scale-dependent bias parameter $b(k,\vbc)$; in \sref{sec:results}, we
present the results of a first demonstration of our methodology and
discuss the findings in
Sections~\ref{sec:discussion}~and~\ref{sec:conclusions}. We present a
comparison of our methodology to some previous works in
Appendix~\ref{sec:comp}. Throughout this paper, we assume a flat
$\Lambda\mathrm{CDM}$ cosmology consistent with the Planck 2018 results
\citep{planckcollaboration2020} with parameters: $\om = 0.314$,
$\ol=0.686$, $\ob=0.049$, $n_s=0.965$, $\sigma_8=0.812$ and
$h=0.673$\footnote{Wherever units are expressed in terms of $h$, it
  can be taken to be this value.}.

\section{Theory}
\label{sec:th}

In this section, we briefly review the relevant theoretical background,
directing the reader to \citet{fialkov2014} and \citet{barkana2016}
for more comprehensive reviews.

In the non-linear regime, the evolution of the density and velocity of
baryons and dark matter is governed by\com{2.3}{:}
\begin{equation}
\begin{array}{l}
\displaystyle\frac{\partial\delta_{\rm c}}{\partial t}+\frac{1}{a} \bm{ v}_{\rm c}\cdot\nabla \delta_{\rm c} = -\frac{1}{a}(1+
 \delta_{\rm c})\nabla\cdot\bm{ v_{\rm c}},\\\\
\displaystyle\frac{\partial\delta_{\rm b}}{\partial t}+\frac{1}{a} \bm{ v}_{\rm b}\cdot\nabla \delta_{\rm b} = -\frac{1}{a}(1+
 \delta_{\rm b})\nabla\cdot\bm{ v_{\rm b}},\\\\
\displaystyle \frac{\partial \bm{ v}_{\rm c}}{\partial t}+\frac{1}{a} (\bm{ v}_{\rm
  c}\cdot\nabla)\bm{ v}_{\rm c} = -\frac{\nabla \Phi}{a}-H\bm{ v}_{\rm c},\\\\
\displaystyle\frac{\partial \bm{ v}_{\rm b}}{\partial t}+\frac{1}{a} (\bm{ v}_{\rm
  b}\cdot\nabla)\bm{ v}_{\rm b} = -\frac{\nabla \Phi}{a}-H\bm{ v}_{\rm
  b}-\com{2.9}{\frac{\nabla p}{a \bar{\rho}_{\rm b}(1+\delta_{\rm b})}},\\\\
 \displaystyle \frac{\nabla^2\Phi}{a^2} = 4\pi G\bar\rho_{\rm m}\delta_{\rm m},
\end{array}
\label{eq:sys1}
\end{equation} 
where $\delta_{\rm b}$ and $\delta_{\rm c}$ are dimensionless perturbations in
baryonic and dark matter densities respectively, $\bm{ v}_{\rm b}$ and
$\bm{ v}_{\rm c}$ are the velocities of \com{2.1}{baryonic} and dark matter
respectively, $a$ is the scale factor, $H \equiv \dot a/a$ is the Hubble
parameter, \com{2.9}{$\Phi$ is the gravitational potential, $p$ is the baryonic pressure, and $\bar{\rho}_{\rm b}$ and $\bar{\rho}_{\rm m}$ are the average densities of baryons and total matter, respectively.}

Following \citet{tseliakhovich2010}, we split the velocities into a
coherent bulk motion, $\bm{v}_{\rm bc}$ (of magnitude $\vbc$), and
random velocity perturbations, $\bm{u}_{\rm b}$ and $\bm{u}_{\rm c}$,
so that in the cold dark matter frame, the velocities can be written as
$\bm{v}_{\rm b} = \bm{v}_{\rm bc}+ \bm{u}_{\rm b}$ and
$\bm{v}_{\rm c} = \bm{u}_{\rm c}$.

\com{2.2}{Though they are second-order terms}, components involving
$\bm{v}_{\rm bc}$ are large for typical values of $\vbc$ at high
redshifts. In addition, these terms become effectively first order on
scales where $\vbc$ is coherent. In this quasi-linear regime,
perturbations in density ($\delta_{\rm b}$ and $\delta_{\rm c}$) and velocities
($\bm{u}_{\rm b}$ and $\bm{u}_{\rm c}$) evolve according to the
following set of equations\com{2.3}{:}
\begin{equation}
\begin{array}{l}
  \displaystyle\frac{\partial\delta_{\rm c}}{\partial t}= -\theta_{\rm c},\\\\
\displaystyle\frac{\partial\delta_{\rm b}}{\partial t} = -\frac{{\rm i}}{a}\bm{v}_{\rm bc}\cdot \bm{k}\delta_{\rm b}-\theta_{\rm b},\\\\
\displaystyle \frac{\partial { \theta}_{\rm c}}{\partial t} = -\frac{3H^2}{2}(\Omega_{\rm
  c}\delta_{\rm c}+\Omega_{\rm b}\delta_{\rm b})-2H\theta_{\rm c},\\\\
\begin{split}
\displaystyle\frac{\partial { \theta}_{\rm b}}{\partial t}=& -\frac{{\rm i}}{a}\bm{v}_{\rm bc}\cdot \bm{
  k}\theta_{\rm b}-\frac{3H^2}{2}(\Omega_{\rm c}\delta_{\rm c}+\Omega_{\rm b}\delta_{\rm
  b})-2H\theta_{\rm b} \\
 & + \com{2.9}{\frac{k_{\rm  B} \bar{T}}{\mu m_{\rm p}}\frac{k^2}{a^2}\left( \delta_{\rm b} +  \delta_T \right)},
\end{split}
\end{array}\label{eq:sys3}
\end{equation} 
where $\theta_i = a^{-1}\nabla \cdot {\bm{u}_i}$ is the velocity divergence in
comoving coordinates \citep[though contrary to
\citealt{tseliakhovich2010}, we work in the rest frame of the dark
matter; see also the appendix of][]{oleary2012}, \com{2.9}{$\delta_T$ is
  the fluctuation in the baryons' temperature, $k_{\rm B}$ is the Boltzmann
  constant, $\mu$ is the mean molecular weight, and $m_{\rm p}$ is the
  mass of the proton.} \com{2.4}{Both the baryon and dark matter
  density parameters $\Omega_{\rm b}$ and $\Omega_{\rm c}$ are functions of
  $t$ (we drop the explicit dependence for clarity)}.

The importance of baryonic pressure for the growth of density modes was
stressed by \cite{naoz2005, naoz2007}, which requires solving an extra
equation to track fluctuations in the temperature $\delta_T$. We follow
\citet{bovy2013} and \citet{ahn2016} in neglecting tracking
fluctuations in photon density and temperature within the evolution
equations, since they are subdominant at most of our scales and
redshifts of interest. We then add the equation for the temperature fluctuations
\begin{equation}
\frac{\partial\delta_T}{\partial t} = \frac{2}{3}\frac{\partial \delta_{\rm b}}{\partial
  t}-\frac{x_{\rm e}(t)}{a^4 t_\gamma} \frac{\bar T_\gamma}{\bar T}\delta_T
\label{eq:t_fluc}
\end{equation}
 to equations~\eqref{eq:sys3}, where
\begin{equation}
t_\gamma^{-1} = \frac{8}{3} \frac{\bar \rho_{\gamma,0} \sigma_{\rm T} c}{m_{\rm e}} = 8.55\times10^{-13}~\mathrm{yr}^{-1}
\end{equation}
and $\bar{T}_\gamma=2.726\ {\mathrm K}/a$ is the mean photon
temperature, $x_{\rm e}(t)$ is the electron fraction out of the total
number density, of gas particles at time $t$, $\bar{T}$ is the mean
gas temperature, \com{2.7, 2.8}{$\bar \rho_{\gamma,0}$ is the mean photon
  energy density at $z=0$, $\sigma_{\rm T}$ is the Thomson scattering
  cross-section for an electron and $m_{\rm e}$ is the mass of an
  electron}. Both $x_{\rm e}(t)$ and $\bar{T}$ are calculated using
\textsc{recfast++} \citep{seager1999,chluba2010, chluba2011}. The
initial conditions for $\delta_T$ are set as in \cite{naoz2005} by
requiring that
$\partial(\delta_T - \delta_{T_\gamma} )/\partial t = 0$ at the initial redshift
$z=1000$, where $\partial \delta_{T_\gamma}/\partial t$ is calculated from \textsc{camb}
\citep{lewis2000}.  The above set of linearised equations can be
solved using a publicly available code, e.g. {\sc cicsass}
\citep{oleary2012}. We reimplement a {\sc python} version of {\sc
  cicsass}, which complements our methodology. \com{2.9}{In the
  current implementation, for simplicity, we ignore the directionality
  of $\bm{v}_{\rm bc}$ when solving this set of equations, instead
  taking
  $\bm{v}_{\rm bc}\cdot \bm{ k}=\vbc k~\mathrm{cos}\theta=\vbc k$
  \rev{}{(i.e. assuming $\bm{v}_{\rm bc}$ is parallel to $\bm{k}$)}. Future
  implementations will take the directionality of $\bm{v}_{\rm bc}$ into account.}
\begin{figure}
  \centering
  \includegraphics[width=\linewidth]{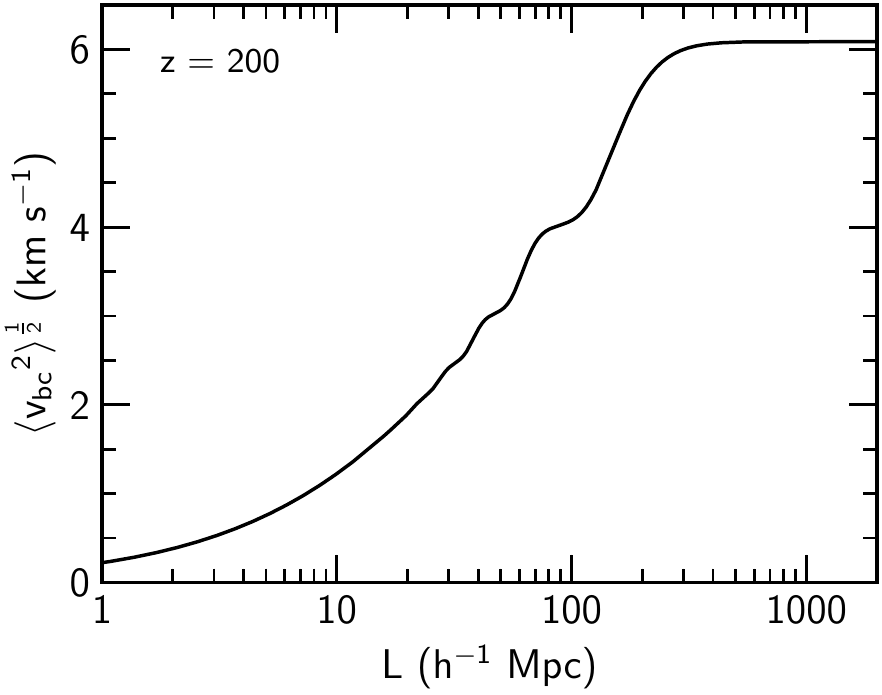}
  \caption{RMS $\vbc$ at $z=200$ as a function of box size, calculated
    by integrating the $\vbc$ power spectrum $\Delta_{{\vbc}}^2$ (computed
    for the cosmological parameters listed in \sref{sec:int}) from
    $2\pi / L$ to $3365~h~{\rm Mpc^{-1}}$.  \com{2.12, 2.13}{The RMS $\vbc$
      converges for $L\gtrsim400~\mpch$, since $\Delta^2_{\vbc}$ drops to zero at
      small-$k$, and so box sizes of this or larger are needed to
      capture all the relevant scales.}}
  \label{fig:rms-vbc}
\end{figure}

\citet{tseliakhovich2010} showed that most of the contributions to the
variance of $\vbc$ come from scales between
$0.005~h~{\rm Mpc^{-1}}$ and $0.5~h~{\rm Mpc^{-1}}$. In a similar
fashion to \citet{pontzen2020}, we can compute the RMS $\vbc$ inside a
box of size $L$ by integrating the power spectrum of $\vbc$
fluctuations from the fundamental mode of the box
$k_{\rm min}=2 \pi/L$ to infinity. The mean square $\vbc$ in a box of
size $L$ is given by
\begin{align}
  \label{eq:rms-vbc}
  \langle\vbc^2\rangle_L =  \int_{2\pi/L}^{k_{\rm max}} {\rm d} k \frac{\Delta_{{\vbc}}^2}{k},
\end{align}
where $\Delta_{{\vbc}}^2$ is the dimensionless power spectrum of the
$\vbc$, taken from {\sc camb} \com{2.11}{and, in theory, the upper limit
$k_{\rm max}$} of the integral in \eref{eq:rms-vbc} should be the
maximum wavenumber of the box, dictated by the number of simulation
elements. In practice, however, any upper limit
$k_{\rm max} \gg 0.5~h~{\rm Mpc^{-1}}$ is sufficient, since the $\vbc$
power spectrum drops off rapidly above this value. \fref{fig:rms-vbc}
shows the RMS $\vbc$, calculated as the square root of
\eref{eq:rms-vbc}, where the oscillatory nature of $\Delta_{\vbc}^2$ at
low-$k$ \citep[cf. fig.~1 in][]{tseliakhovich2010} is clearly visible.
From \fref{fig:rms-vbc}, we can see that even in a box size of
$100~\mpch$, we do not capture all of the scales relevant to
$\vbc$. The curve only begins to plateau around $\sim400~\mpch$, so using
a box size smaller than this means that we may miss out on some of the
effect, for example by not sampling extreme values of
$\vbc$. Simultaneously simulating this large-scale box and the very
high-resolution zoom region needed to observe the effect would be
computationally infeasible. In \sref{sec:ics}, we discuss our solution
to this problem.

\section{Methods}
\label{sec:method}

\subsection{Simulations}
\label{sec:sim}

We follow the evolution of dark matter, gas, and stars in the
cosmological context using {\sc ramses}\footnote{The version used here
  is commit {\tt aa56bc01} from the \texttt{master} branch. Note that
  older versions of \textsc{ramses} may not use separate fields for
  dark matter and baryon velocities by default.}, which employs a
second-order Godunov method to solve the equations of
hydrodynamics. Gas states are computed at cell interfaces using the
Harten-Lax-van Leer-contact Riemann solver, with a MinMod slope
limiter. Dark matter and stars are modelled as a collisionless
$N$-body system, described by the Vlasov-Poisson equations.  Grid
refinement is performed whenever a cell contains more than eight
high-resolution dark matter particles, or has the equivalent amount of
baryonic mass scaled by $\Omega_{\rm b}/ \Omega_{\rm m}$. We allow the
AMR grid to refine from the coarsest level $\ell_{\rm min} = 8$ to the
finest level $\ell_{\rm max} = 23$, but in practice grid hold-back within
\textsc{ramses} means that the finest level reached is $\ell=21$,
corresponding to a maximum comoving resolution of
$47.7~\pch$\footnote{\com{3.1}{Releasing higher levels of AMR grids at
    specified steps in scale factor (`grid hold-back') is a technique
    employed in \ramses{} to ensure that the physical (as opposed to
    comoving) resolution remains roughly constant over an entire
    simulation, which is desirable for e.g.\@ ensuring that cells at
    high redshift do not over-refine and end up containing too little
    mass to form stars. \citet{snaith2018} performed a detailed study
    \rev{}{of} the effects of grid hold-back on simulation properties,
    finding, among other results, that the sudden release of
    high-resolution grids can lead to spikes in the star formation
    rate. \rev{3}{Our simulation reaches its highest refinement level
      $\ell=21$ before any star formation occurs, and so is not impacted
      by these grid hold-back effects on star formation}}}.

Star formation is allowed whenever the gas density of a cell is
greater than $n_\star=1~\mathrm{cm^{-3}}$ in units of the number density
of hydrogen atoms and when the local overdensity is greater than
$200\rho_{\rm cr}$, where the latter condition prevents spurious star
formation at extremely \com{3.2}{high redshift}. We impose a polytropic
temperature function with index $g_\star=2$ and
$T_0=1050\, \mathrm{K}$, which ensures that the Jeans length is always
resolved by at least eight cells. We do not rigorously calibrate the
star formation parameters to reproduce any stellar mass-halo mass
relation, since we are interested only in the differences between
simulations. Star particles, which represent a population of stars,
form with a mass of $108.0~\msolh$. Supernova feedback is included
using the kinetic feedback model of \citet{dubois2008}, with a mass
fraction $\eta_{\rm SN} = 0.1$ and a metal yield of 0.1. We allow gas
cooling and follow the advection of metals. We do not include
molecular hydrogen in this simulation so, to attempt to compensate for
this missing cooling channel, we initialise the zoom region with a
metallicity of $Z=10^{-3}~{\rm Z_\odot}$, where ${\rm Z_\odot} = 0.02$ in
\textsc{ramses}.

\subsection{Initial conditions}
\label{sec:ics}

As described in \sref{sec:th}, large box sizes of $\ga 400~\mpch$ are
required in order to capture all of the scales pertaining to
$\vbc$. By performing calibration runs, we found that very high
resolution (a cell size of $\Delta x \la 2~\kpch$) is needed in the ICs in
order to properly resolve the effect. To this end, we employ `zoom'
initial conditions (ICs), generating density and velocity fields at
$z_{\rm ini}=200$ first in a $400~\mpch$ box using \textsc{music}
\citep{hahn2011}. The ICs are refined from the base level
$\ell_{\rm min} = 10$ ($1$,\@$024^3$) up to $\ell = 18$
($262$,\@$144^3$ effective) in a cube of side length $543~\kpch$ at
the finest level. \rev{A}{Such extremely high resolution is required
  because $\vbc$ suppresses structure formation on very small
  scales. We found that the resolution we used in this work is the
  minimum necessary to observe the effect of $\vbc$, and using lower
  resolution largely misses the effect.} Since the zoom region is
very small compared to the box size, we use extra padding between zoom
levels, increasing the number of padding cells on each side for each
dimension from the typical value of $4$ to $32$. We use transfer
functions from \textsc{camb} \citep{lewis2000}, which gives distinct
density and velocity fields for the baryons and dark matter.

\begin{table}
  \centering
  \caption{The sets of ICs used for the main zoom simulation. The
    columns list the name of each case, which velocity fields are used
    for the baryons ($\bm{v}_{\rm b}$), whether the baryon fields have
    been modified and the magnitude of $\vbc$ at $z=1000$
    ($v_{\rm bc, rec}$), and $z=200$ ($v_{\rm bc, ini}$) in $\kms$. If
    a value is present for $v_{\rm bc, ini}$, but not
    $v_{\rm bc, rec}$, then $\vbc$ is only included from the start
    time of the simulation. }
\label{tab:ics}
\begin{tabular}{lcccc}
\hline
Case & $\bm{v}_{\rm b}$  & Modified? & $v_{\rm bc, rec}$ & $v_{\rm bc, ini}$ ($\kms$)\\
\hline
\novbc{} & $\bm{v}_{\rm c}$ & no & 0.0 & 0.0\\
\unbiased{} & $\bm{v}_{\rm b}$  & no & 0.0 & 20.09\\
\biased{} & $\bm{v}_{\rm b}$  & yes & 100.07 & 20.09\\
\hline
\end{tabular}
\end{table}

In order to make the simulation tractable, we extract a $100~\mpch$
($\ell_{\rm min}=8$, $256^3$) base grid from the $400~\mpch$
($\ell_{\rm min}=10$, $1$,\@$024^3$) box and use this as our coarsest
level, meaning that the maximum refinement level in the zoom region
also drops two levels from $\ell=18$ to $\ell=16$ ($65$,\@$536^3$
effective). In principle, this methodological choice could introduce
some error around the edges of the box due to using periodic boundary
conditions with non-periodic ICs, though in practice we expect the
impact of this to be negligible since we are concerned with a
sub-$\mpch$ region in the centre of the box. \rev{5}{Additionally,
  these errors will be common between runs, so their effect will wash
  out when comparing between runs.}

\tref{tab:ics} details the sets of ICs used in this work. We selected
a region for zoom-in with $v_{\rm bc, ini}=20.09~\kms$ at $z=200$,
corresponding to $v_{\rm bc,rec}=100.07~\kms$, or
$\sim3.3\sigma_{\vbc}$, at recombination. The \novbc{} case is often used in
cosmological simulations, for example when using transfer functions
that do not have separate amplitudes for the baryon and dark matter
velocity fields (in fact, it is the default behaviour for older
versions of {\sc ramses}, where the dark matter velocity field is used
to initialise both the dark matter and baryon velocities). The
\unbiased{} case is where the simulation is initialised using separate
transfer functions for the baryon and dark matter velocity fields,
such as by generating ICs using {\sc music} with transfer functions
from {\sc camb}. In this case, $\vbc$ is included from the start time
of the simulation $z_{\rm ini}$, but the effect of $\vbc$ on density
and velocity perturbations between recombination and $z_{\rm ini}$ is
missed. In the final, and most realistic, case, \biased{}, we include
the contributions from $\vbc$ across all $z$ by computing a bias
factor which is applied to the ICs. The methodology for computing the
bias factor is detailed in \sref{sec:bias}.

\subsection{Bias factor}
\label{sec:bias}
\begin{figure}
  \centering
  \includegraphics[width=\columnwidth]{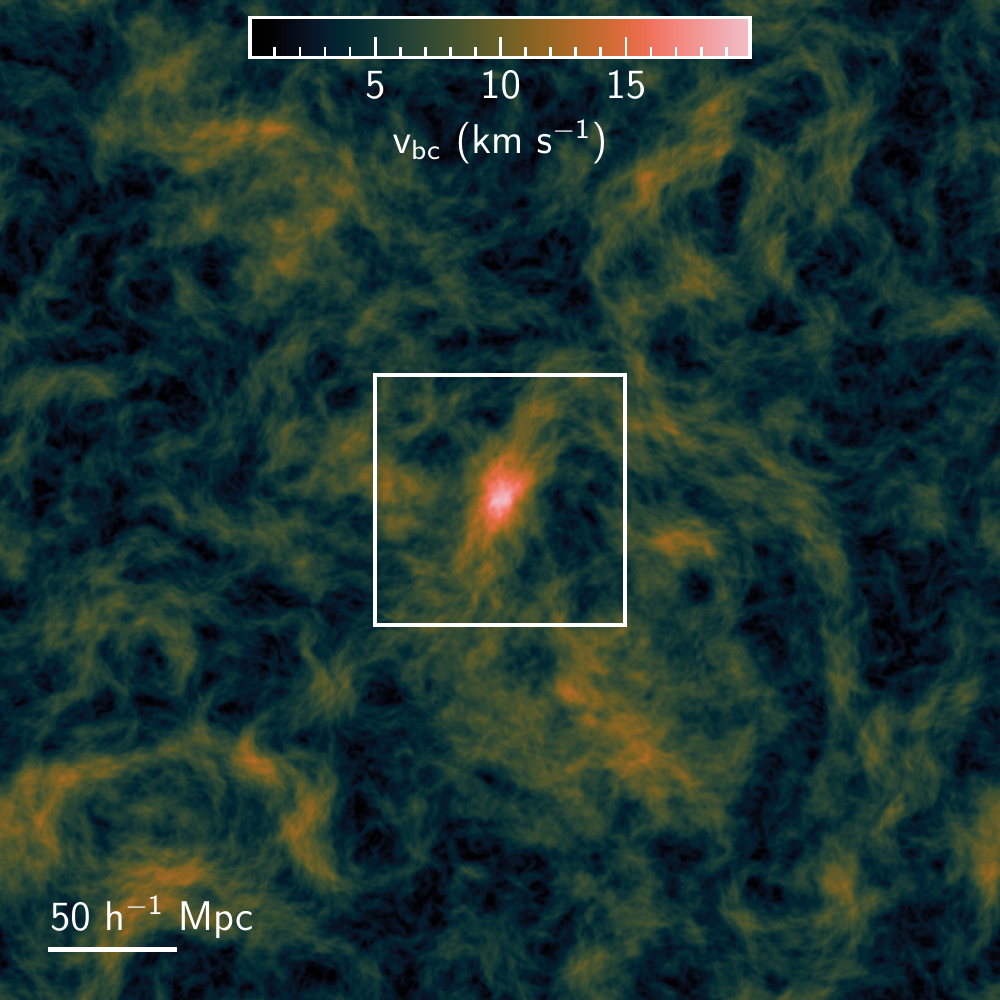}
  \caption{A slice through the large-scale
    $\vbc = |\bm{v}_{\rm b} - \bm{v}_{\rm c}|$ field in the full
    $400~\mpch$ initial conditions at $z_{\rm ini}=200$. Each pixel
    corresponds to a cell width of $0.39~\mpch$ and the slice
      has a thickness of one cell width. Also shown (white square) is
    the position of the extracted $100~\mpch$ subbox, centred on the
    peak $\vbc$ in the ICs. The zoom region is located in the centre
    of the subbox. The colour shows the magnitude of the $\vbc$ at
    $z_{\rm ini}=200$, where light pink is high $\vbc$ and dark blue
    is low $\vbc$.\label{fig:vslc}}
\end{figure}
Using transfer functions that have distinct amplitudes for the baryon
and dark matter velocity fluctuations naturally yields the $\vbc$
field at the start time of the simulation $z_{\rm ini}$. First, we
interpolate the dark matter particle velocities onto the same grid as
the baryons, then take the difference of these two fields to calculate
the magnitude as
$\vbc = \left | \bm{v}_{\rm b} - \bm{v}_{\rm c} \right |$. A
$0.39~\mpch$ thick slice through the resultant $\vbc$ field is shown
in \fref{fig:vslc}.

\begin{figure}
  \centering
  \includegraphics[width=\columnwidth]{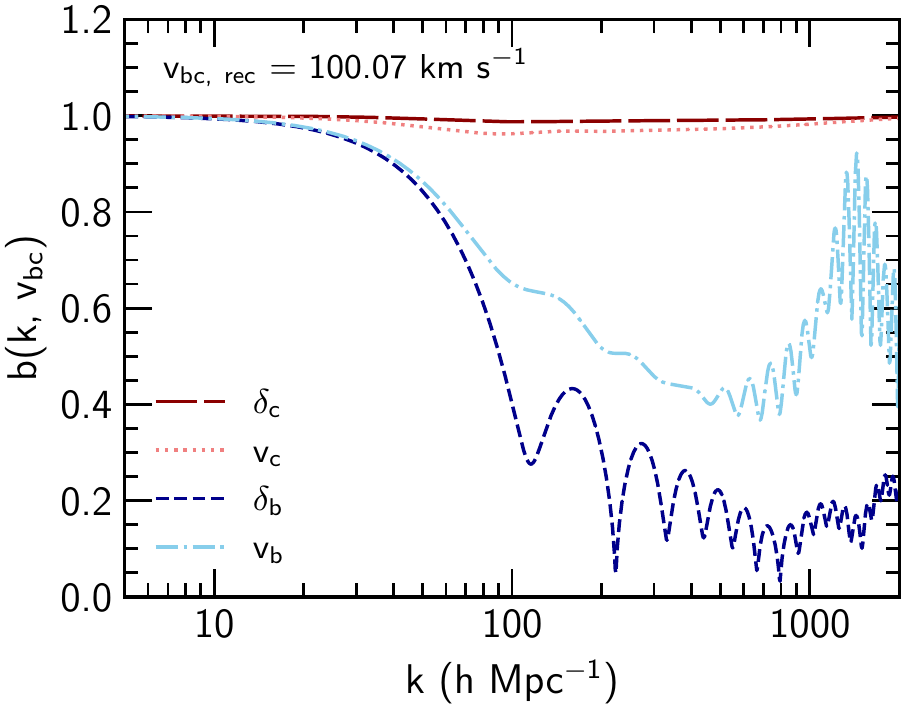}
  \caption{The bias factors $b(k\,\vbc)$ at $z_{\rm ini}=200$ for the
    average $\vbc$ in the zoom region, scaled back to its value at
    recombination $v_{\rm bc, rec}=100.07~\kms$, which is a
    $\sim3.3\sigma_{\vbc}$ value. We show $b(k, \vbc)$ for baryon and dark
    matter overdensities and peculiar velocities. These are the bias
    factors that are applied to the perturbations in the zoom region.
    Note how the perturbations in the baryon overdensity (dark blue,
    short-dashed) are strongly suppressed for $k>40~\impch$, while the
    perturbations in the dark matter overdensity (dark red,
    long-dashed) are largely unaffected, to the few per~cent
    level. Note that we show $b(k, \vbc)$ for the peculiar velocities
    for the full range of $k$, but the velocity field is dominated by
    large-scale (small-$k$) modes. Therefore, $b(k, \vbc)$ will have
    little if any impact on the small scales (large-$k$) \com{3.3}{of the
    velocity field.} \label{fig:bk}}
\end{figure}

\begin{figure*}
  \centering
  \includegraphics[height=.9\textheight]{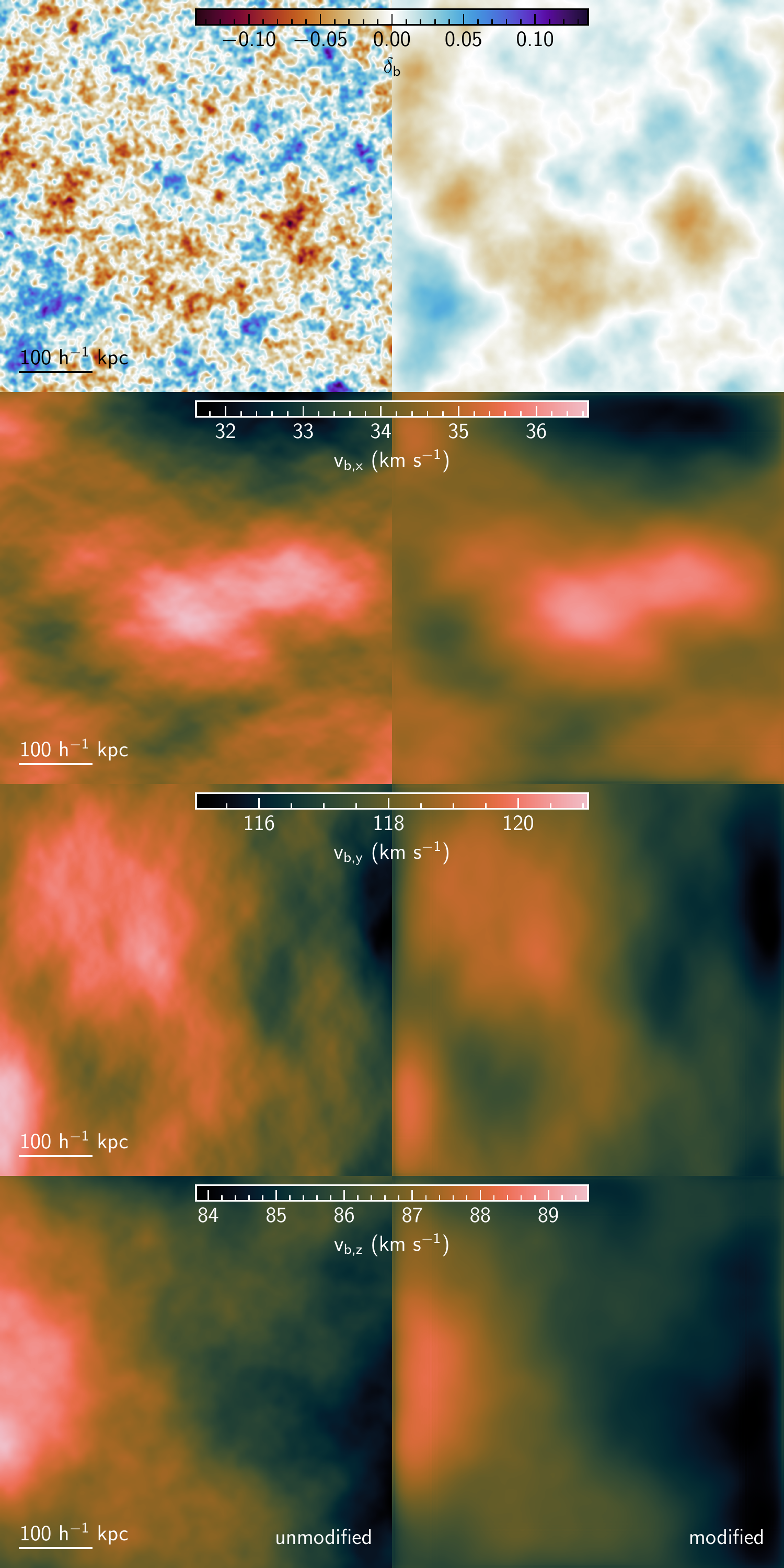}      
  \caption{Slices of the unmodified (left column) and modified
    (\rev{A}{through convolution with the bias factor, as in
      \eref{eq:bias-conv}}, right column) baryon overdensity (top
    row), peculiar velocity in the $x$-direction (second row),
    $y$-direction (third row) and $z$-direction (bottom row) in the
    high-resolution zoom region, of side length $543~\kpch$. Each
    pixel corresponds to a cell width of $1.53~\kpch$ and the slice
    has a thickness of one cell width. The effect of applying
    $b(k, \vbc)$, \com{3.9}{defined in \eref{eq:bias_factor}}, can be
    clearly seen in the baryon overdensity, in that it washes out the
    small-scale fluctuations. The effect is less pronounced in the
    peculiar velocities, which are dominated by \com{3.7}{large-scale}
    modes.}
  \label{fig:bias-slc}
\end{figure*}

With the $\vbc$ field in hand, we split our ICs into cubic patches,
aiming for a patch extent of $0.5~\mpch$, though the actual extent
depends upon how many patches can be fit in each level of the
\textsc{grafic} files. The size of these patches is chosen to be
smaller than the scale over which $\vbc$ is coherent
\citep{tseliakhovich2010}. Within each patch, the average value of
$\vbc$ is calculated and used as $\vbc$ in
equations~\eqref{eq:sys3}. The initial values for
equations~\eqref{eq:sys3} are set using the transfer functions from
{\sc camb} at $z=1000$, and the equations are integrated from $z=1000$
to $z=200$ using the {\sc lsoda} ordinary differential equation
solver. Equations~\eqref{eq:sys3} are solved for the average patch
value of $\vbc$ and also for $\vbc=0~\kms$, which yields power spectra
for the baryon perturbations both with and without $\vbc$. We use
these power spectra to calculate a `bias' factor at $z_{\rm ini}=200$
that depends both upon scale $k$ and the magnitude of the relative
velocity $\vbc$
\begin{equation}
  b(k, \vbc) = \left [\frac{P(k, \vbc)}{P(k, \vbc=0)} \right]^{\frac{1}{2}},
  \label{eq:bias_factor}
\end{equation}
where the square root arises from $P\propto\left |\delta^2 \right|$. In
\fref{fig:bk}, we show the bias factor for the baryon and dark matter
densities ($\delta_{\rm b}$ and $\delta_{\rm c}$) and velocities
($v_{\rm b}$ and $v_{\rm c}$), computed for the average $\vbc$ in our
zoom region. The strongest suppression is seen in the baryons and in
particular the baryon density, while the dark matter is hardly
affected. \aut{3}{We do not expect the oscillatory features in
  $b(k,v_{\rm bc})$ at the very small scales to have much, if any,
  impact since the power spectrum of fluctuations in the baryon
  density contrast begins to fall rapidly for $k\gtrsim 300~\impch$, while
  for the velocity most of the power is at much larger
  scales. \citet{ali-haimoud2014} also found oscillatory features in
  the small-scale baryon perturbations, and we have checked that we
  find similar oscillations for typical values of $\vbc$ and also find
  that increasing the magnitude of $\vbc$ increases the frequency of
  oscillations for the larger-scale ($\gtrsim 100~\impch$) modes too.} \rev{4}{For a detailed study into the origin of these small-scale oscillations, we defer the reader to \citet{ali-haimoud2014}.}  This
factor is then convolved with the Fourier transform of the
corresponding patch of baryon overdensity
\begin{equation}
  \hat{\delta}_{\rm b}(k, \vbc) = b(k, \vbc) \cdot \delta_{\rm b}(k)
  \label{eq:bias-conv}
\end{equation}
to give individual patches of biased overdensity
$\hat{\delta}_{\rm b}$, which are then stitched together to generate the
\biased{} set of ICs. In this way, the bias factor compensates for the
suppression of baryon perturbations between $z=1000$ and $z_{\rm ini}$
that is missing if $\vbc$ is included only from $z_{\rm ini}$. We
only modify the baryons, since as discussed earlier, they are much
more strongly affected than the dark matter, as can be seen from
\fref{fig:bk}.

We deal with the peculiar velocity field for the baryons in a similar
way, by first converting the velocity divergence to peculiar
velocities as $\bm{v}_{\rm b}(k) = -{\rm i}a\bm{k}\theta_{\rm
  b}(k)/k^2$. Note again that we do not include directionality when
solving the evolution equations, and therefore, the bias factor is
applied to each direction of $\bm{v}_{\rm b}$ equally. In reality,
there would be preferential directions for the bias factor, depending
on the direction of $\bm{v}_{\rm bc}$, but we defer that
implementation \rev{A}{to future work}.

\fref{fig:bias-slc} shows a $1.53~\kpch$ thick slice through the
highest resolution level of the zoom ICs directly from {\sc music}
(`unmodified', left column) and after the bias factor $b(k, \vbc)$ has
been applied (`modified', right column). For $\delta_{\rm b}$ (top row),
the unmodified ICs contain a lot of small-scale structure, which is
almost totally washed out after applying $b(k, \vbc)$. Most of what
remains is in the form of lower amplitude, larger scale
fluctuations. For $v_{{\rm b},i}$ (bottom rows)\footnote{Note that we
  show the $v_{\rm b}$ for each direction for completeness, but the
  effect is independent of direction in our methodology.}, there is
less small-scale structure to begin with, since the peculiar velocity
fields are dominated by large scales. The effect of $b(k,\vbc)$ on
$v_{{\rm b},i}$ is therefore much less striking than on the
$\delta_{\rm b}$, with the main effect being smoothing and a slight
reduction in amplitude.

\subsection{Haloes \label{sec:haloes}}
After the ICs have been correctly initialised with $\vbc$, we can
characterise the effect of $\vbc$ on structure formation, principally
by exploring how haloes are affected. Haloes are identified using
\textsc{ahf} \citep{gill2004, knollmann2009}, which supports
multi-resolution datasets and calculates baryonic properties of
haloes. In order to use {\sc ahf} with a \textsc{ramses} dataset, we
use the supplied {\sc ramses2gadget} tool to convert leaf cells of the
AMR hierarchy into gas pseudoparticles placed at the centre of each
cell. \com{3.5}{This conversion is a standard recipe in the analysis
  of AMR datasets \citep[e.g.][]{wu2015} and allows us to conveniently
  assign gas to haloes.} We ignore the internal energy of the gas and
do not allow \textsc{ahf} to perform any unbinding, as this has been
shown to remove most of the gas from subhaloes in a manner that is
dependent upon the choice of halo finder \citep{knebe2013}. We define
the halo overdensity with respect to the critical density
$\rho_{\rm cr}$, such that the average density inside the halo is
$200\rho_{\rm cr}$.

Guided by the resolution study of \citet{naoz2007}, we perform our analysis
on haloes that have $\ge500$ particles, as they found that this is the
level at which the baryon fraction is resolved with a scatter of
$\sim20~{\rm per~cent}$ when compared to higher resolution simulations.

We only use haloes comprised entirely of high-resolution dark matter
particles, since contaminant low-resolution particles can disrupt the
dynamics of haloes \citep{onorbe2014}. Due to the small size of the
zoom region, it is often the case that haloes which were initially
solely composed of high-resolution particles can become contaminated
by low-resolution particles as the simulation progresses. If
contaminated haloes are removed at each timestep, then haloes can
`disappear' if they become contaminated between one timestep and the
next. To counter this effect, we generate merger trees using {\sc
  consistent-trees} \citep{behroozi2013a} and only keep haloes that
have never been contaminated at any point in the
simulation. \com{3.6, 4.3, 4.4}{Then, when exploring the effect on
  halo properties (such as when looking at their gas content), we go
  one step further and match haloes between the simulations.} We do
this using {\sc ahf}'s {\sc MergerTree} tool to correlate the particle
IDs between each run. This allows us to isolate the effect on
identical haloes, as opposed to also capturing the impact on the
global population of haloes. \com{3.6, 4.3, 4.4}{The difficulties in
  exploring the impact on the global population of haloes
  (e.g. through the cumulative number of haloes) \rev{}{are} explored in
  \sref{sec:hmf}.}

\section{Results}
\label{sec:results}

\com{3.6, 4.3, 4.4}{
\subsection{Halo abundances}
\label{sec:hmf}
Throughout this section, we calculate the cumulative number of haloes $N(>M)$
as the number of haloes with a total mass greater than $M$ and
estimate the Poisson uncertainty on $N(>M)$ for each case as
$\sqrt{N(>M)}$. To begin with, we ignore the conditions described in
\sref{sec:haloes}, instead considering the cumulative number of all haloes
formed in the simulation $N_{\rm all} (>M)$, shown in
\fref{fig:hmf-all}. Aside from a slight suppression in
$N_{\rm all}(>M)$ for the \biased{} case below $10^6~\msolh$ at
$z=14.2$, the \unbiased{} and \biased{} cases are consistent both with
each other and with the \novbc{} case at the $1\sigma$ level. Analysing the
haloes in this fashion (i.e. by performing no cleaning on the
catalogue) gives a picture of the global impact of $\vbc$ on halo
abundance, however this picture is inaccurate because a significant
fraction of the haloes are contaminated by, or formed entirely of,
lower-resolution particles, affecting the accuracy of their properties
(see \sref{sec:haloes} for a discussion of contamination).

\label{sec:hmf-all}
\begin{figure}
  \centering
  \includegraphics[width=\linewidth]{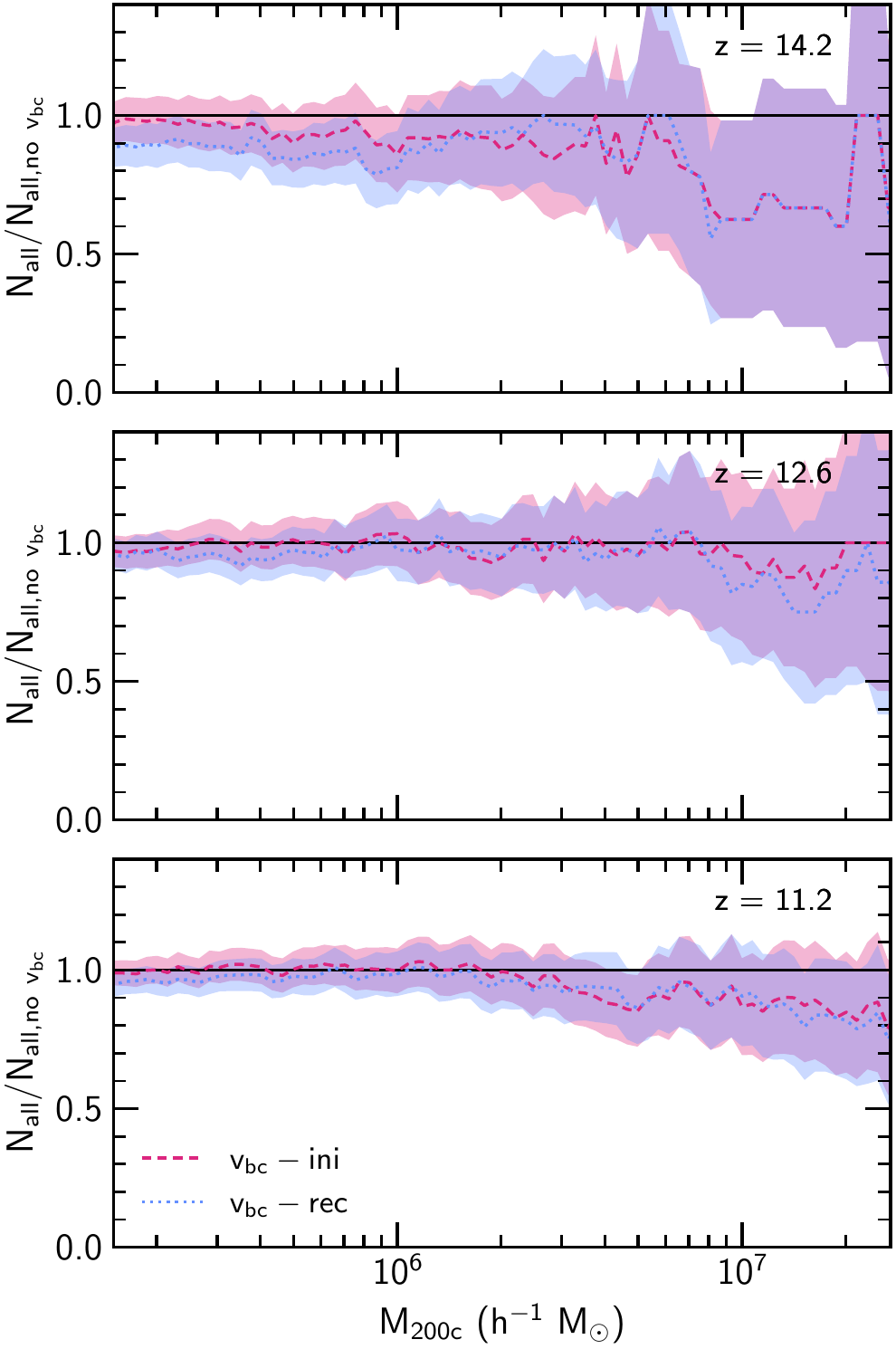}
  \com{3.6, 4.3, 4.4}{\caption{Ratio of the cumulative number of
      haloes $N_{\rm all}(>M)$ in the \unbiased{} (pink short-dashed)
      and the \biased{} (blue dotted) cases to the \novbc{} case at
      $z=14.2$ (top), $12.6$ (middle), and $11.2$ (bottom). All the
      haloes found by {\sc ahf} are included in $N_{\rm all}(>M)$. The
      shaded regions \com{4.2}{indicate} the $1\sigma$ Poisson uncertainty
      on the ratio. $N_{\rm all}(>M)$ is broadly the same between all
      three sets of simulations, except for a suppression in the
      number of haloes with $M<10^6~\msolh$ at $z=14.2$ in the
      \biased{} case compared to the \novbc{}
      case.} \label{fig:hmf-all}}
\end{figure}

If we now apply the conditions described in \sref{sec:haloes}, namely
that we do not include in our analysis any haloes that are
contaminated at any point in the simulation, we are left with a
reduced catalogue. In \fref{fig:hmf-clean}, we show the cumulative
number of haloes in this cleaned catalogue $N_{\rm clean}(>M)$, which
shows a much more striking suppression of $N_{\rm clean}(>M)$ for the
\biased{} case and an apparent increase in the abundance of low-mass
($M<10^6~\msolh$) haloes for the \unbiased{} case at $z=11.2$ (though
still consistent with no difference to the \novbc{} case at the
$1\sigma$ level).

\label{sec:hmf-clean}
\begin{figure}
  \centering
  \includegraphics[width=\linewidth]{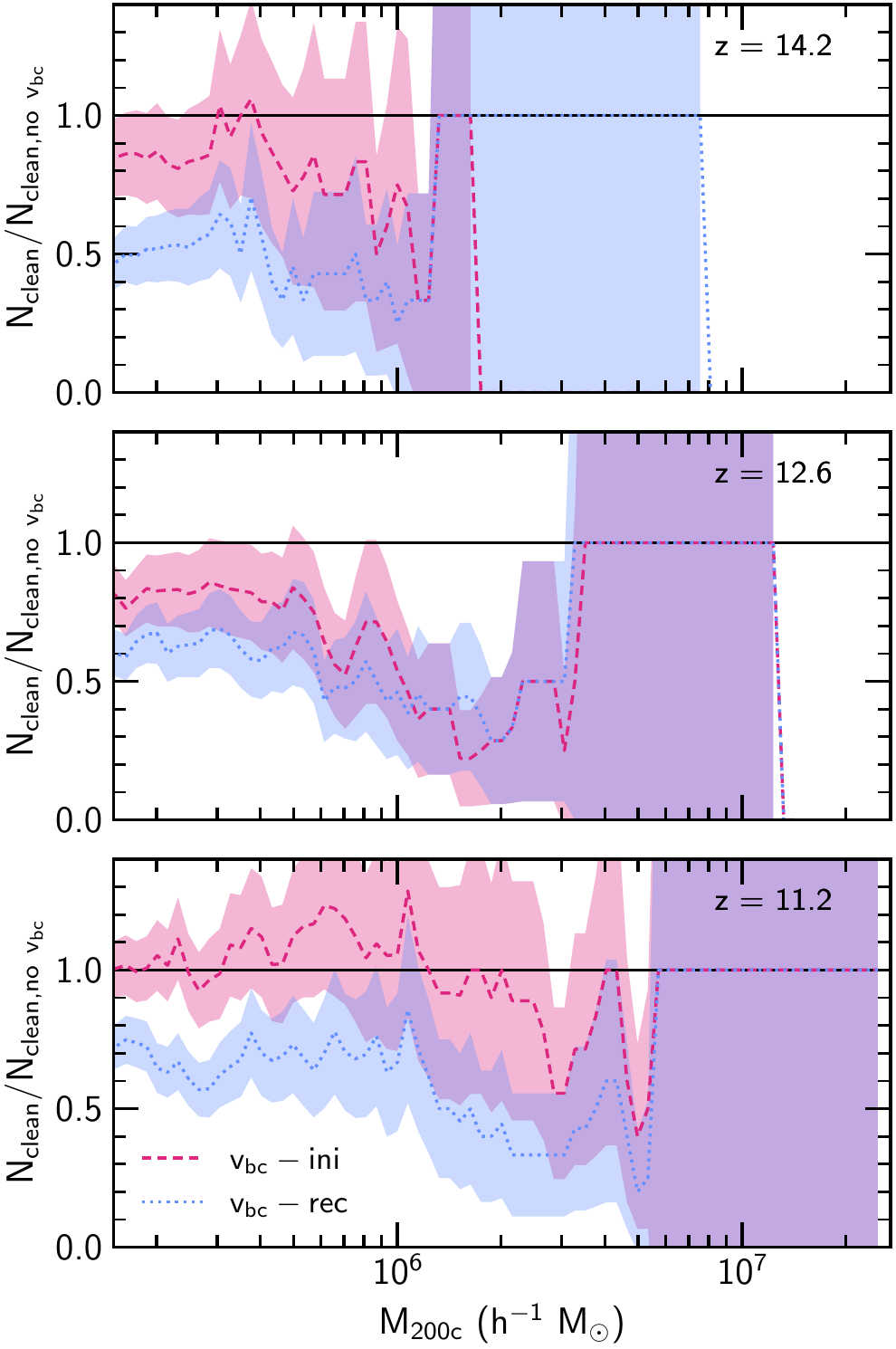}
  \com{3.6, 4.3, 4.4}{\caption{As in \fref{fig:hmf-all}, but this time $N_{\rm clean}(>M)$
    includes only haloes that have been selected by the process
    detailed in \sref{sec:haloes}. Selecting never-contaminated haloes
    in this fashion leads to the removal of an unequal and
    inconsistent number of haloes between each run---this leads to the
    odd behaviour of the \unbiased{} curve, which jumps from a
    suppression to a boost in the number of haloes with
    $M<2\times10^6~\msolh$ between $z=12.6$ and $11.2$.}\label{fig:hmf-clean}}
\end{figure}

To highlight the effect of the cleaning procedure we also show the raw
cumulative number of haloes at $z=14.2$ and $z=11.2$ in
\fref{fig:hmf-raw}. The impact of the cleaning process described in
\sref{sec:haloes} can be clearly seen in the bottom panel of
\fref{fig:hmf-raw}, which shows the ratio of the cleaned catalogue to
the full catalogue $N_{\rm clean}(>M)/N_{\rm all}(>M)$. Comparing the
different runs, it can be clearly seen that a different fraction of
haloes is removed between each run at each of the times shown.

\label{sec:hmf-raw}
\begin{figure}
  \centering
  \includegraphics[width=\linewidth]{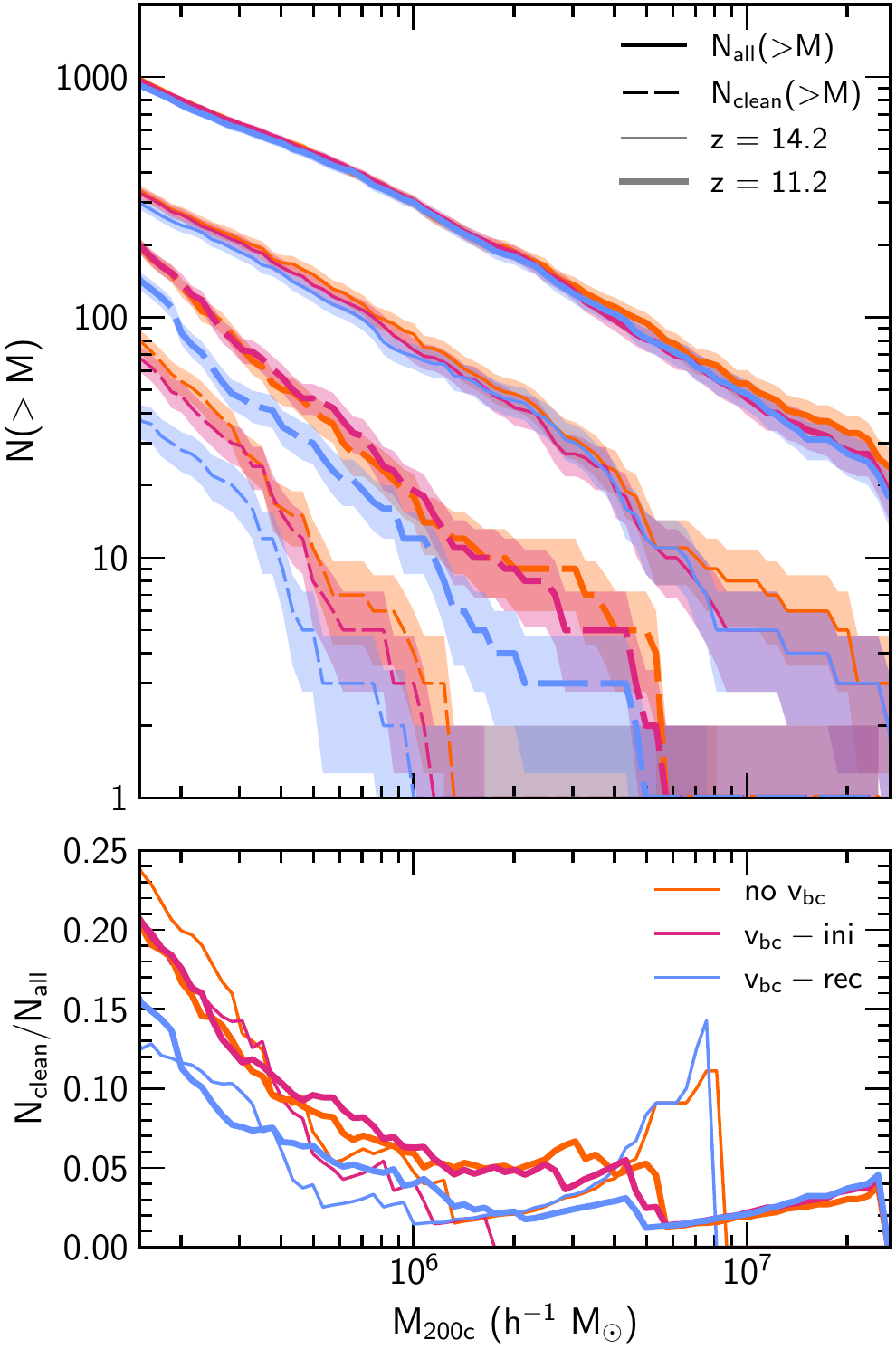}
  \com{3.6, 4.3, 4.4}{\caption{Cumulative number of haloes (top panel)
      for the raw catalogue $N_{\rm all}(>M)$ (solid) and the
      catalogue cleaned according to the prescription in
      \sref{sec:haloes} $N_{\rm clean}(>M)$ (long-dashed) and the
      \rev{}{ratio} of $N_{\rm clean}(>M)$ to $N_{\rm all}(>M)$ (bottom panel)
      at each $z$. We show $N(>M)$ at $z=14.2$ (thin) and $11.2$
      (thick). From the top panel, it can be seen that
      $N_{\rm all}(>M)$ is very similar between all runs at each $z$
      shown, except for a slight suppression in the abundance of
      haloes in the \biased{} case at $z=14.2$. The huge reduction in
      the number of haloes after cleaning the catalogue is abundantly
      clear, as is the difference in fraction of haloes removed
      between each run.}  \label{fig:hmf-raw}}
\end{figure}
}

\subsection{Baryon fraction} 

\begin{figure}
  \centering
  \includegraphics[width=\linewidth]{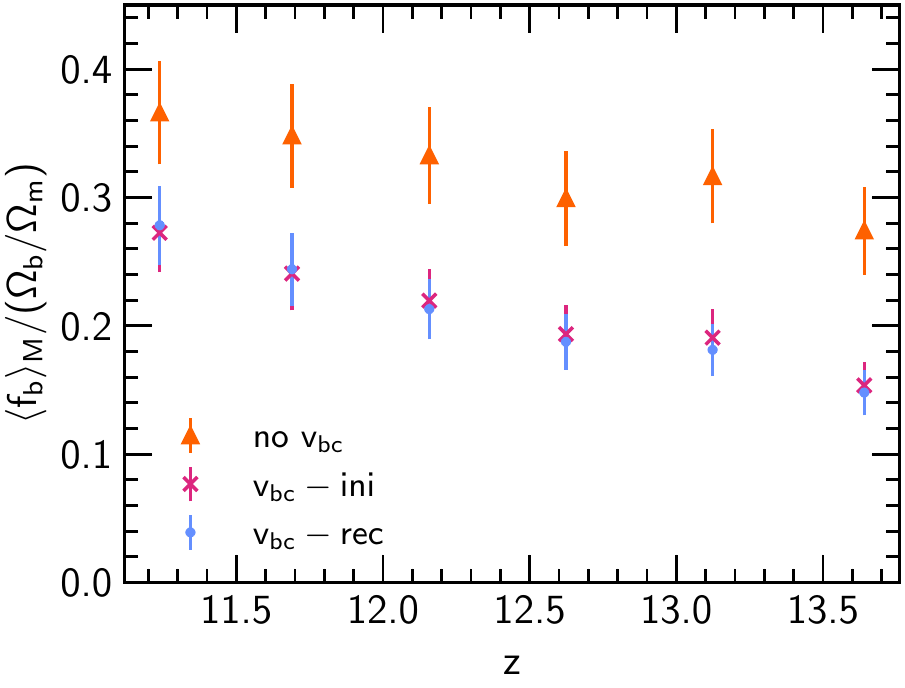}
  \caption{Mass-weighted average baryon fraction $\fb$ as a function
    of redshift $z$, normalised to the cosmic mean
    $\Omega_{\rm b} / \Omega_{\rm m}$. The average is shown for all
    $z$ where more than $30$ haloes have formed. The errorbars show
    the $1\sigma$ standard deviation as calculated using
    \eref{eq:mw-std}. \com{2.13}{We find a suppression in both the
      $\unbiased{}$ and $\biased{}$ cases compared to the \novbc{}
      run, though between the two runs with $\vbc$ there is little
      difference.}}
  \label{fig:mw-fb}
\end{figure}

We allow star formation in these runs, so the total baryon fraction of
a given halo is defined as
\begin{equation}
  \fb = \frac{M_{\rm g} + M_{\star}}{M_{\rm d} + M_{\rm g} + M_{\star}},
\end{equation}
where $M_{\rm g}$ is the gas, $M_{\star}$ the stellar, and $M_{\rm d}$ the
dark matter mass in each halo. We upweight the best resolved (i.e.\
most massive) haloes by calculating the mass-weighted average baryon
fraction as
\begin{align}
  \langle\fb\rangle_M=\frac{\sum_{i}f_{{\rm b}, i}M_i}{\sum_i M_i}
\end{align}
and the associated mass-weighted standard deviation as
\begin{align}
  \label{eq:mw-std}
  \sigma_M = \left (\frac{\sum_if^2_{{\rm b}, i}M_i}{\sum_i M_i} - \langle\fb\rangle_M^2 \right ) ^{\frac{1}{2}},
\end{align}
where the sum is over all haloes that satisfy the conditions in
\sref{sec:haloes}. \fref{fig:mw-fb} shows $\langle\fb\rangle_M$ and associated
$1\sigma$ errorbars as function of $z$. We show each $z$ where
  $\ge30$ haloes have formed that satisfy the criteria in
  \sref{sec:haloes}, starting from $z=13.6$ where we are able to match
  $38$ haloes between the three cases. The gas fraction is suppressed
at all $z$ for the \unbiased{} and \biased{} cases compared to the
\novbc{} case. At earlier $z$, the suppression is stronger, though
even by the final snapshot at $z=11.2$, $\langle\fb\rangle_M$ for both the
\biased{} and \unbiased{} cases are not within $1\sigma$ of the \novbc{}
case. Notably, at all $z$, $\langle\fb\rangle_M$ in \unbiased{} and \biased{}
cases are almost indistinguishable from, and certainly consistent
with, one another.

\subsection{Star formation}
\begin{figure}
  \centering
  \includegraphics[width=\linewidth]{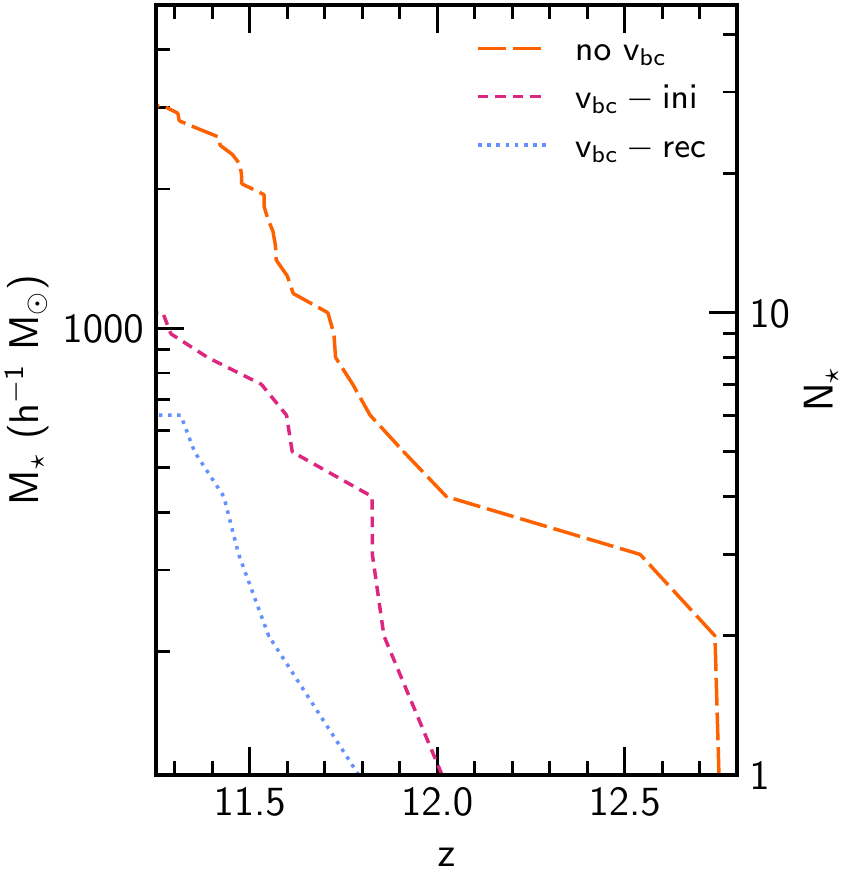}
  \caption{Cumulative stellar mass $M_\star$ formed as function of
    redshift $z$. Also shown is the corresponding number of stellar
    particles $N_\star$. \com{2.13}{Despite the small number of stars
      formed in this simulation, there is a clear hierarchy in how
      many stars each forms, with the \novbc{} run forming the most,
      followed by $\unbiased{}$ and then $\biased{}$. There is also a
      clear delay in the onset of star formation which follows a
      similar trend, with the $\biased{}$ run starting to form stars
      the latest.}}
  \label{fig:mstar}
\end{figure}

In \fref{fig:mstar}, we show the cumulative $M_\star$ formed in the
simulation, not accounting for mass loss due to supernovae, and the
corresponding number of stellar particles $N_\star$, which each have a
mass of $108.0~\msolh$. In each case, all of the star particles
  in the simulation formed inside a single halo. In total, $29$ star
particles formed by $z=11.2$ in the \novbc{} case, $10$ in the
\unbiased{} case, and $7$ in the \biased{} case. This hierarchy
persists across all $z$, with more star particles having formed
in the \novbc{} case than in the \unbiased{} case and fewer still in
the \biased{} case.

\com{4.6}{In the following, all times quoted in $\myr$ are measured
  relative to the Big Bang. To quantify how $\vbc$ impacts star
  formation in our simulations, we can calculate the delay in
  formation time of the $n$th star $t_n$ for each run with $\vbc$
  compared to the \novbc{} run $t_{n,\, \text{no}~\vbc}$ as
  $\Delta t_n = t_n - t_{n,\, \text{no}~\vbc}$. For the \unbiased{} case,
  the smallest (largest) delay of $8.6~\myr$ ($35.5~\myr$) occurs for
  the formation of the fourth (second) star particle, while for the
  \biased{} case we find a delay of $23.4~\myr$ ($49.4~\myr$) for the
  sixth (second) star particle to form.
  
  Star formation in {\sc ramses} \rev{A}{(like in other simulation
    codes)} is a stochastic process, hence the large fluctuations in
  delay times. We can reduce the impact this stochasticity has on our
  results by averaging over delay times, as opposed to considering the
  delay times for individual star particles. We find a mean delay time
  of $\langle\Delta t\rangle=19.4~\myr$ and $34.9~\myr$ for the \unbiased{} and
  \biased{} cases, respectively.}

\section{Discussion}
\label{sec:discussion}

\com{3.6, 4.3, 4.4}{To understand the perceived uptick in \unbiased{} haloes in
\fref{fig:hmf-clean}, we need to explore the raw cumulative number of haloes,
shown in \fref{fig:hmf-raw} at $z=14.2$ and $z=11.2$. The key point is
that while only keeping haloes that are never contaminated ensures
that haloes do not disappear between timesteps of the same run, it
also means that different numbers of haloes will be removed between
different runs at any given timestep. This discrepancy in the
number of haloes removed is responsible for the apparent low-mass
boost in the \unbiased{} case, because over the mass range
$3\times10^5<M/\msolh<10^6$ fewer haloes are removed in the \unbiased{}
case than in the \novbc{} case. Since the cumulative numbers of haloes are
already very similar, removing a smaller number of haloes in the
\unbiased{} case manifests itself as a boost relative to the \novbc{}
case, when in fact it is simply an artefact of the cleaning procedure.

Given the difficulties in extracting a sample of haloes that is both
free from contamination and comparable between runs, we will not draw
any quantitative conclusions on the impact of $\vbc$ on global
properties like the number of haloes formed. We retain this discussion
of the well-known impact of contamination and, crucially, the
importance of verifying any mitigation techniques as it may prove
helpful for other works employing zoom simulations.}

Including $\vbc$ also significantly affects the baryon fraction $\fb$,
where we see that the mass-weighted baryon fraction
$\langle\fb\rangle_M$ is suppressed at all redshifts in both cases, with the
suppression stronger at higher redshift. Even by $z=11.2$, the
\unbiased{} and \biased{} cases are still not in agreement with the
\novbc{} case, though the difference between the two populations has
decreased. Again, this is likely due to the decay in the magnitude of
$\vbc$, which allows the haloes to accrete more gas. Interestingly,
$\langle\fb\rangle_M$ is almost indistinguishable between the \unbiased{} and
\biased{} cases, \com{4.5}{suggesting that including $\vbc$ from
  $z_{\rm ini}$ is sufficient to observe its impact on halo baryon
  fraction, though a larger sample is required to confirm this.} The
suppression in $\langle\fb\rangle_M$ when $\vbc$ is included is in qualitative
agreement with previous studies.

This decrement in baryon fraction for the \unbiased{} and \biased{}
cases is reflected in the cumulative stellar mass formed, as fewer
star particles formed in both cases than in the \novbc{} case. Not
only do they form fewer star particles, they also start forming star
particles later since the effect of $\vbc$ is to wash out the peaks
(and troughs) in the baryon density contrast, meaning that it takes
longer for gas to reach the densities required for star formation.
The mean delay for forming star particles is $19.4~\myr$ for the
\unbiased{} case and $34.9~\myr$ for the \biased{} case.  From
\citet{schaerer2002}, we find that these delays are all of the order
of the lifetime of a $9~\msol$ first-generation Population (Pop) III
star, which has a lifetime of $20.02~\myr$ \citep[table 3
in][]{schaerer2002}. More massive Pop III stars have even shorter
lifetimes, for example a $120~\msol$ Pop III star lives for only
$2.52~\myr$. Pop III stars form from initially pristine gas, and their
death pollutes their immediate surroundings with metals, introducing
new cooling channels into the high-redshift Universe. Any delay in
this introduction of metals will delay the transition between Pop III
to Pop II (i.e.\ from metal-enriched gas), which can, for example,
affect the $21~{\rm cm}$ signal \citep[][]{magg2022}. In our case,
though we do not form Pop III stars, chemical enrichment is still
vitally important for star formation to get properly underway,
particularly as all of the star particles form in the same halo.

Despite there being almost no difference in $\langle\fb\rangle_M$ between the \unbiased{}
and \biased{} cases at most redshifts, there is a clear hierarchy in
the amount of stars formed---\novbc{} forms the most, \unbiased{}
forms fewer, and \biased{} forms the least---albeit on the order of a
few star particles. This effect is expected, since the bias factor
washes out baryonic density peaks, and there are slightly more haloes
(i.e.\ star formation locations) present in the \unbiased{} case than
in the \biased{} case.

\section{Conclusions}
\label{sec:conclusions}

We have performed the first cosmological zoom simulations that
self-consistently sample the relative baryon-dark matter velocity
$\vbc$ from a large $400~\mpch$ box. This relative velocity naturally
arises when simulations are initialised using transfer functions that
have separate amplitudes for the baryon and dark matter velocities,
and we have shown that a box roughly as large as this is
required to properly sample all of the scales associated with the
relative velocity. However, solely initialising simulations in this
manner misses out on the effect of the relative velocities from
$z=1000$ to the start time of the simulation, $z_{\rm ini}$. We
developed a methodology that compensates for the effect of $\vbc$
on baryon density and velocity perturbations by computing a `bias'
factor $b(k, \vbc)$, which is convolved with the ICs. We verified that
our methodology performs as expected by comparing to previous works
(see Appendix~\ref{sec:comp}).

As a first demonstration of our methodology, we applied it to an
extremely high-resolution zoom region in a $100~\mpch$ subbox,
extracted from the main $400~\mpch$ box. The zoom region is centred on
the region with the largest relative velocity in the $400~\mpch$ box,
which has an RMS value of $\vbc=100.07~\kms$ at $z=1000$,
corresponding to $\sim3.3\sigma_{\vbc}$. \aut{8}{We find qualitative agreement
  with previous works, namely a reduction in the halo baryon fraction,
  a delay in the onset of star formation at high redshift, and a
  suppression of the final stellar mass.} The strength of the effect
decreases with redshift, but the two simulations still exhibit some
differences by $z=11.2$. We find that the delay in the onset of star
formation is of the order of the lifetime of a $\sim9~\msol$ Pop III
star. We also test the effect of incorporating the bias factor by
running a simulation that includes the relative velocity from the
start time of the simulation only. In this case, we find that more
stars are formed when compared to the simulation that includes the
bias factor, but there is almost no change in the average baryon
fraction, except at the earliest redshift. \com{3.6, 4.3, 4.4}{Due to
  the small size of our zoom region, the vast \rev{}{majority} of haloes in
  our simulation are contaminated with low-resolution particles, and
  we are thus unable to draw any robust conclusions regarding the halo
  mass function.}

Our code for producing these compensated ICs is publicly
available\footnote{\url{https://github.com/lconaboy/drft}}, and we
hope will be of use for studying this effect in the full cosmological
context.

\section*{Acknowledgements}

\aut{4}{We thank the anonymous referee for constructive feedback which
  improved this paper. LC thanks Antony Lewis and Joakim Rosdahl for
  helpful discussions.} \rev{A}{LC was supported by an STFC
  studentship and acknowledges the indirect support of a Royal Society
  Dorothy Hodgkin Fellowship and a Royal Society Enhancement
  Award. ITI was supported by the Science and Technology Facilities
  Council [grant numbers ST/I000976/1 and ST/T000473/1] and the
  Southeast Physics Network.} AF is supported by the Royal Society
University Research Fellowship URF/R1/180523. This work used the
DiRAC@Durham facility managed by the Institute for Computational
Cosmology on behalf of the STFC DiRAC HPC Facility
(\url{www.dirac.ac.uk}). The equipment was funded by BEIS capital
funding via STFC capital grants ST/K00042X/1, ST/P002293/1,
ST/R002371/1 and ST/S002502/1, Durham University and STFC operations
grant ST/R000832/1. DiRAC is part of the National
e-Infrastructure. The authors gratefully acknowledge the Gauss Centre
for Supercomputing e.V. (\url{www.gauss-centre.eu}) for funding this
project by providing computing time through the John von Neumann
Institute for Computing (NIC) on the GCS Supercomputer JUWELS at
J\"ulich Supercomputing Centre (JSC). We acknowledge that the results
of this research have been achieved using the PRACE Research
Infrastructure resource Galileo based in Italy at CINECA. We
acknowledge PRACE for awarding us access to Beskow/Dardel hosted by
the PDC Center for High Performance Computing, KTH Royal Institute of
Technology, Sweden. This work made use of the following software
packages: {\sc numpy} \citep{harris2020}; {\sc matplotlib}
\citep{hunter2007}; {\sc scipy} \citep{virtanen2020}; {\sc yt}
\citep{turk2011}; {\sc ytree} \citep{smith2019}; {\sc hmf}
\citep{murray2013} and {\sc cmasher} \citep{vandervelden2020}.

\section*{Data Availability}
\label{sec:data_av}
The code for computing and applying the bias factor $b(k, \vbc)$ is
available at \url{https://github.com/lconaboy/drft}. The data
underlying this article will be shared on reasonable request to the
corresponding author.



\bibliographystyle{mnras}
\bibliography{refs} 




\appendix

\section{Comparison to previous works}
\label{sec:comp}
We run a series of test simulations set up as in \citet{naoz2012,
  naoz2013}\footnote{\aut{7}{A later study performed using the
    moving-mesh code \textsc{arepo} found better agreement with the
    analytical gas fraction relation \citep{popa2016, chiou2018}. For
    the comparison in this section, \citet{naoz2012, naoz2013} are
    sufficient.}}. The specific case shown here has a base resolution
$\ell_{\rm min}=9$ ($512^3$ dark matter particles and, initially, cells),
in a $471.1~\kpch$ periodic box. The simulations in \citet{naoz2012,
  naoz2013} were performed using the SPH code {\sc gadget2}
\citep{springel2005}, while we use the AMR code {\sc ramses}. We allow
the AMR grid to refine freely up to $\ell_{\rm max}=14$, corresponding to
a maximum comoving resolution of $28.9~\pch$, comparable to the
gravitational softening length of $45.8~\pch$ comoving used in
\citet{naoz2012}. We use our fiducial cosmology (\sref{sec:sim}),
whereas \citet{naoz2012} used a cosmology consistent with
\citet{komatsu2009} with parameters: $\om = 0.28$, $\ol=0.72$,
$\ob=0.046$, $h=0.7$\footnote{When lengths and masses are quoted in
  units of $h$, we use $h=0.673$ from our choice of cosmology.}  and a
boosted $\sigma_8=1.4$. We adopt the boosted $\sigma_8=1.4$ as used in the
original studies. We initialise the simulations at
$z_{\rm ini}=200$\footnote{The \citet{naoz2012} simulations are
  actually initialised at $z_{\rm ini}=199$ but the $\vbc$ would only
  have decreased by $0.5~{\rm per~cent}$ in this time, so we ignore
  this difference.} both with and without a relative velocity of
$1.7\sigma_{\vbc}=10~\text{km~s}^{-1}$ at $z_{\rm ini}$. We also compute
and apply the bias factor $b(k, \vbc)$ to the baryonic component of
the ICs, while the \citet{naoz2012} simulations are initialised by
computing transfer functions which explicitly include the effect of
the $\vbc$. Following \citet{naoz2012}, we set both of the velocity
fields to that of the dark matter and apply the $\vbc$ to the
$x$-component of the baryon velocity as
\begin{align}
  \begin{pmatrix}
    v_{\mathrm{b}, x} \\
    v_{\mathrm{b}, y} \\
    v_{\mathrm{b}, z} 
  \end{pmatrix}
  =
  \begin{pmatrix}
    v_{\mathrm{c}, x} + \vbc \\
    v_{\mathrm{c}, y} \\
    v_{\mathrm{c}, z}
  \end{pmatrix}.
\end{align}
Including the $\vbc$ in this manner is justified by the box size being
much smaller than the coherence scale of the $\vbc$. We do not allow
star formation in these runs. Haloes are identified as described in
\sref{sec:haloes}, noting that although \citet{naoz2012} define their
halo overdensity with respect to the background matter density
$\bar{\rho}_{\rm m}$, we are working in the period of matter domination
\com{6.1}{so} the difference will be small. \aut{2}{While the final
  halo masses in \citet{naoz2012} are defined using a spherical
  overdensity method, the initial halo finding is done using a
  friend-of-friends method, which could be susceptible to
  `overlinking' \citep[e.g.][]{davis1985}, whereby disparate groups of
  particles are spuriously connected by a diffuse particle
  bridge. Their redefinition of the halo mass using the spherical
  overdensity method will go some way to alleviating the impact (if
  any) of overlinking.}

To calculate the effect of the $\vbc$, we compare to
simulations without $\vbc$, where the velocity field of the baryons is
equal to that of the dark matter. In order to quantify this effect, we
calculate the fractional difference of a quantity $A$ as
\begin{equation}
  \Delta_A=\frac{A_{\vbc} - A_{{\rm no}~\vbc}}{A_{{\rm no}~\vbc}}.
  \label{eq:dec}
\end{equation}

\begin{figure}
  \centering
  \includegraphics[width=\columnwidth]{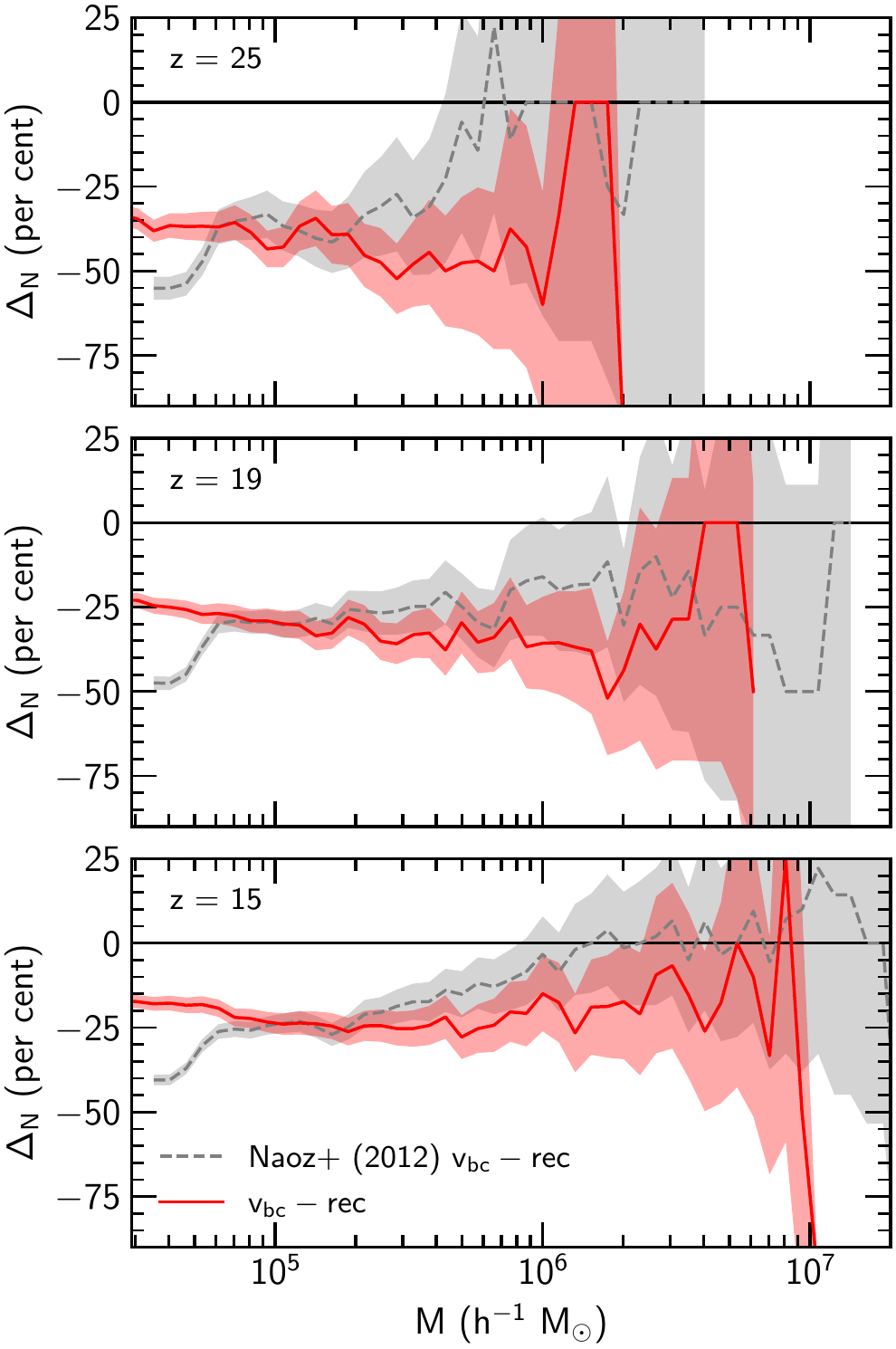}
  \caption{The fractional difference $\Delta_N$ in halo mass function
    $N(>M)$ for the $v_{\rm bc, ini}=10~\kms$ case, calculated with
    \eref{eq:dec}. We show $\Delta_N$ at $z=25$ (top), $19$ (centre), and
    $z=15$ (bottom) for our simulations (red solid) and for the
    \citet{naoz2012} work (grey dashed). \com{2.13, 6.5}{The shaded regions
      indicate the $1\sigma$ Poisson uncertainty on the ratio. At most
      masses we are consistent with \citet{naoz2012}, though there is
      some disagreement for $M<6\times10^4~\msolh$, which could arise because of the friends-of-friends algorithm used to initially find haloes in \citet{naoz2012}.\label{fig:naoz-ngm}}}
\end{figure}
First, we look at the effect on the cumulative halo mass function
$N(>M)$, as in \citet{naoz2012}. \fref{fig:naoz-ngm} shows the
decrement in $N(>M)$ for the case with $\vbc$ compared to the case
without, both for our simulations and for the \citet{naoz2012} run. We
see qualitatively similar behaviour, observing a decrement between $\sim
0~{\rm per~cent}$ and $50~{\rm per~cent}$ at all redshifts shown and for almost all masses.

\begin{figure}
  \centering
  \includegraphics[width=\columnwidth]{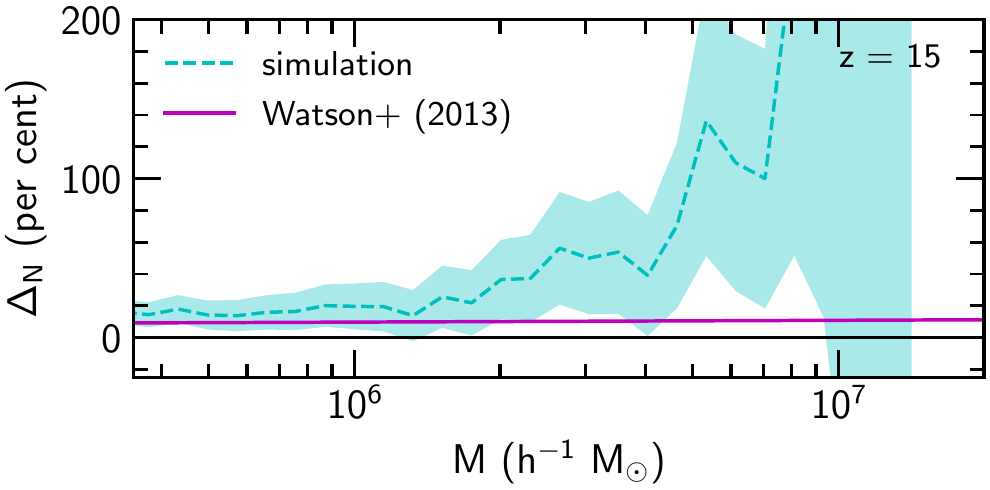}
  \caption{Comparison of the fractional difference $\Delta_N$ in halo mass
    function $N(>M)$ between our simulations and the simulations from
    \citet{naoz2012} (cyan short-dashed) and between the
    \citet{watson2013} curve, which is fit to $N$-body simulations,
    for the cosmology used in our work and the one used in
    \citet{naoz2012} (magenta solid). The fractional difference is
    calculated as $\Delta_N=(N_1-N_0)/N_0$, where $N_0$ are the data
    corresponding to our work and $N_1$ to \citet{naoz2012}, so
    $\Delta_N>0$ means there are more haloes in
    \citet{naoz2012}. \aut{1}{Neither run includes $\vbc$. The shaded
      cyan region indicates the combined 1$\sigma$ Poisson uncertainty on
      $N(>M)$ from the simulations.}\com{2.13, 6.4}{For
      $M \lesssim 4\times 10^6~\msolh$ the difference in $N(>M)$ between the
      simulations is mostly consistent with the expected difference
      due to cosmology and halo mass definitions (i.e. the difference
      between the \citealt{watson2013} curves). Potential sources of
      the high-mass ($M\gtrsim4\times 10^6~\msolh$) discrepancy are discussed in
      the text.
    }}
  \label{fig:naoz-ngm-diff}
\end{figure}
However, the overall shape of our $\Delta_N$ is slightly different to
\citet{naoz2012}; we match well below $\sim3\times10^5~\msolh$ but show more
relative suppression above this mass. This discrepancy is due, at
least in part, to the different simulation codes used and the
different white noise fields in the initial conditions. One further
significant source of difference is the cosmologies
used. \fref{fig:naoz-ngm-diff} shows the difference expected at $z=15$
by comparing the analytic \citet{watson2013} $N(>M)$ mass functions
(magenta solid). From this, we would expect the \citet{naoz2012}
simulation to have $\sim9~{\rm per~cent}$ more haloes with
$M>3\times10^5~\msolh$. This increase in the number of haloes increases
with mass and for $M>1 \times 10^7~\msolh$, we would expect
$\sim 11~{\rm per~cent}$ more haloes in the \citet{naoz2012}
simulation. Indeed this is borne out by the simulations (cyan dashed),
which show that the \citet{naoz2012} simulations do produce more
haloes at all masses. At higher masses, $\Delta_N$ diverges as the absolute
number of haloes becomes small. \aut{2}{As discussed earlier,
  \citet{naoz2012} use the friends-of-friends groups as a starting
  point, which could be a source of the high-mass discrepancy in
  \fref{fig:naoz-ngm-diff}.}

\begin{figure*}
  \centering
  \includegraphics[width=\linewidth]{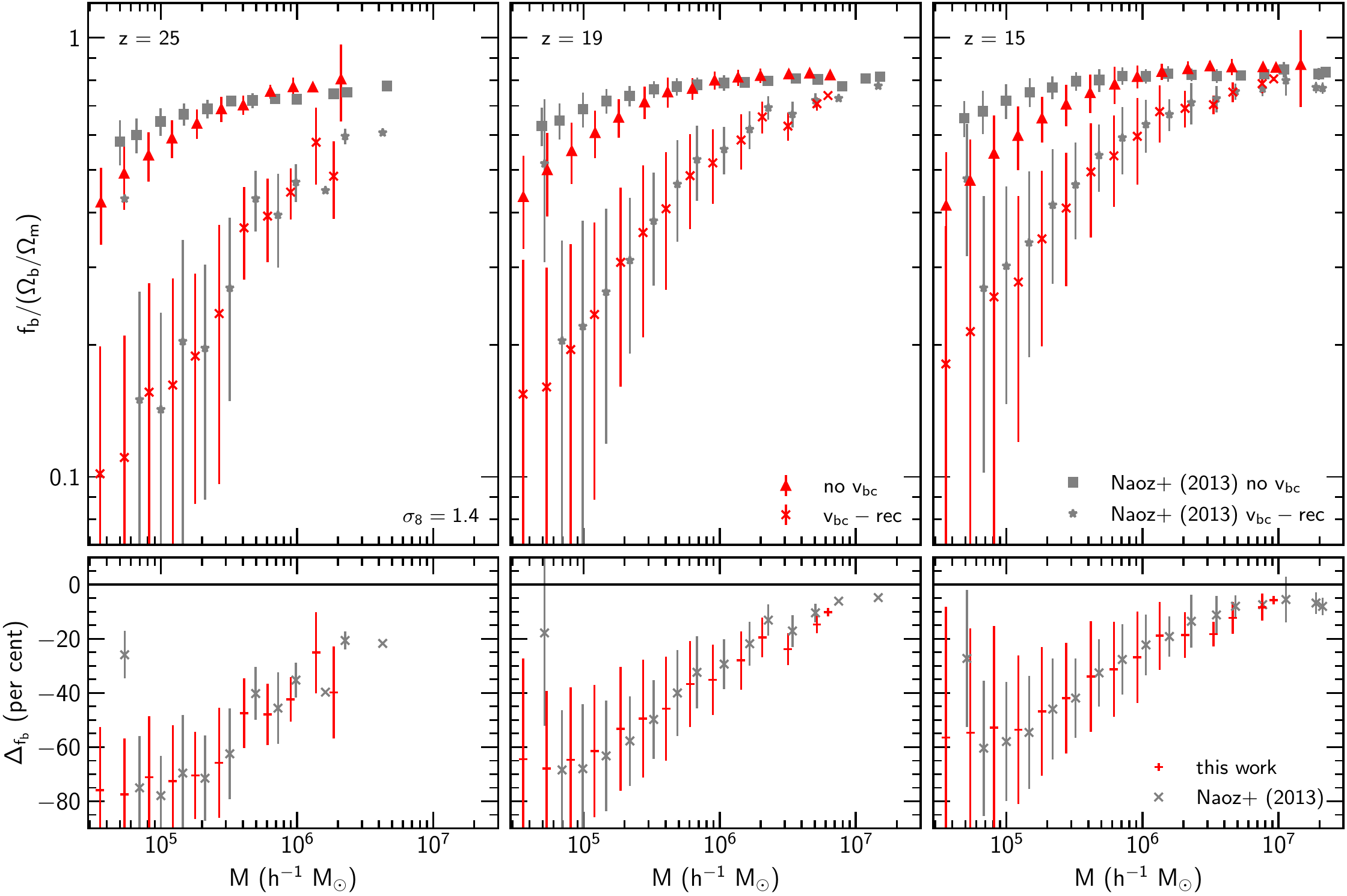}
  \caption{Binned baryon fraction $\fb$ (top panels) and relative
    fractional difference $\Delta_{\fb}$ between each run with $\vbc$ to
    the run without, calculated with \eref{eq:dec} (bottom panels). We
    show data from our simulations (red triangles, points and plusses)
    and from the \citet{naoz2013} work (grey squares, stars and
    crosses). The panels show $z=25$ (left), $19$ (centre) and $15$
    (right). \aut{1}{The errorbars indicate the $1\sigma$ standard deviation in each mass bin (top panels) and the combined $1\sigma$ uncertainty (bottom panels).} \com{2.13, 6.3}{We find excellent agreement in the relative difference
    between our work and \citet{naoz2013}, and broad agreement in the absolute value of $\fb$. Some slight disagreement in the absolute value could be due to the different choice of simulation methodology, since we employ an Eulerian code where \citet{naoz2013} use a Lagrangian code. \label{fig:naoz_fb_bin}}}
\end{figure*}

Next, we turn our attention to the gas fraction of haloes, as studied
in \citet{naoz2013}. Since we do not include star formation in these
runs, the baryon fraction is simply the halo gas mass divided by
the total halo mass
\begin{equation}
  f_{\rm b} = \frac{M_{\rm g}}{M_{\rm g} + M_{\rm d}}.
  \label{eq:fbnos}
\end{equation}
\fref{fig:naoz_fb_bin} shows the binned gas fractions (top panel)
for our and for the \citet{naoz2013} simulations, each normalised to
the cosmic mean $\ob/\om$ for the appropriate
cosmology, and the decrement (bottom panel) as defined in
\eref{eq:dec}. We take the midpoint of the mass bin to be the mean
of all the mass values in that bin. The binned gas fractions for
\citet{naoz2013} are slightly higher than in this work, though they
exhibit roughly the same mass dependence. The agreement between the
two simulations for the decrement is striking -- they have an extremely
similar mass dependence. There is some difference in the binned baryon
fractions, in particular we find slightly more suppression at lower
masses. This is likely due to differences in code used since, as
mentioned previously, \citet{naoz2013} used {\sc gadget2}
\citep{springel2005}, where we use {\sc ramses}. There are well documented
differences between Lagrangian (e.g.\ SPH) and Eulerian (e.g.\
AMR) codes \citep[e.g.][]{agertz2007}, and indeed it has been shown
that numerical diffusion due to baryon-grid relative velocities can
artificially smooth densities in Eulerian codes
\citep{pontzen2020}. In any case, we are not interested in comparing
the merits of different codes, so by calculating the difference
between the runs with and without $\vbc$, we can remove artefacts due
to the choice of code.


\bsp	
\label{lastpage}
\end{document}